\documentclass[Times1COL]{WileyNJDv5} %STIX1COL,STIX2COL,STIXSMALL

\articletype{Article Type}%

\received{Date Month Year}
\revised{Date Month Year}
\accepted{Date Month Year}
\journal{Journal}
\volume{00}
\copyyear{2023}
\startpage{1}

\usepackage[nocompress]{cite}
\usepackage{amsmath,amssymb,commath}
\usepackage{enumitem,hyperref,color}
\usepackage{multirow,graphicx,subfigure}
\usepackage{float}
\usepackage[numbers,super,sort&compress]{NJDnatbib}

\newcommand{\X}{\pmb{X}}
\newcommand{\x}{\pmb{x}}
\newcommand{\Y}{\pmb{Y}}
\newcommand{\W}{\pmb{W}}
\def\E{\mbox{\rm E}}
\def\var{\mbox{\rm Var}}
\def\cov{\mbox{\rm Cov}}
\def\RSS{\mbox{\rm RSS}}
\def\MISE{\mbox{\rm MISE}}

\begin{document}

\title{Nonparametric Regression and Error Covariance Function Estimation - Beyond Short-Range Dependence}

\author[1]{Sisheng Liu}

\author[2]{Xiaoli Kong}

\authormark{Liu and Kong}
\titlemark{PLEASE INSERT YOUR ARTICLE TITLE HERE}

\address[1]{\orgdiv{MOE-LCSM, School of Mathematics and Statistics}, \orgname{Hunan Normal University}, \orgaddress{\state{Changsha}, \country{China}}}

\address[2]{\orgdiv{Department of Mathematics}, \orgname{Wayne State University}, \orgaddress{\state{Detroit, MI}, \country{USA}}}

\corres{Corresponding author Sisheng Liu, MOE-LCSM, School of Mathematics and Statistics, Hunan Normal University, Changsha, Hunan 410081, P. R. China. \email{ssl1989@hunnu.edu.cn}}

\presentaddress{ssl1989@hunnu.edu.cn}

%\fundingInfo{Text}
%\JELinfo{ejlje}

\abstract[Abstract]{	In nonparametric regression analysis, errors are possibly correlated in practice, and neglecting error correlation can undermine most bandwidth selection methods. When no prior knowledge or parametric form of the correlation structure is available in the random design setting, this issue has primarily been studied in the context of short-range dependent errors. When the data exhibits correlations that decay much more slowly, we introduce a special class of kernel functions and propose a procedure for selecting bandwidth in kernel-based nonparametric regression, using local linear regression as an example. Additionally, we provide a nonparametric estimate of the error covariance function, supported by theoretical results. Our simulations demonstrate significant improvements in estimating the nonparametric regression and error covariance functions, particularly in scenarios beyond short-range dependence. The practical application of our procedure is illustrated through the analysis of three datasets: cardiovascular disease mortality, life expectancy, and colon and rectum cancer mortality in the Southeastern United States.}


\maketitle

\renewcommand\thefootnote{}
\footnotetext{\textbf{Abbreviations:} ANA, anti-nuclear antibodies; APC, antigen-presenting cells; IRF, interferon regulatory factor.}

\renewcommand\thefootnote{\fnsymbol{footnote}}
\setcounter{footnote}{1}

\section{Introduction}\label{sec1}

Assume that we observe random samples $(\X_i, Y_i)$ for $i=1,\ldots,n$, where $Y_i$ takes values in $\mathbb{R}$ and $\X_i=(X_{i1},\ldots,X_{iD})^T$ takes values in $\mathcal{X}\subset\mathbb{R}^D$ with a common probability density $f$. The density $f$ is compactly supported, bounded and continuous with $f(\x)>0$ for all $\x\in\mathcal{X}$. 
Consider the model:
\begin{equation}
	Y_i = \mu(\X_i) +\varepsilon_i, \  \text{ for } i=1,..,n, \label{Sec1_model}
\end{equation}
where $\mu$ is an unknown smooth (at least twice continuously differentiable) regression function and the $\varepsilon_i$'s are unobserved random errors such that 
\begin{equation*}
	\E(\varepsilon_i\vert\X_i) = 0, \  \cov(\varepsilon_i,\varepsilon_j \vert \X_i, \X_j) = \sigma^2 \rho_n(\Vert \X_i - \X_j \Vert),\ \text{for } i,j\in\{1,\ldots,n\},
\end{equation*}
with $\sigma^2<\infty$ and $\rho_n$ a stationary correlation function satisfying $\rho_n(0)=1$ and $\abs{\rho_n(\Vert \x_i - \x_j \Vert)} \leq 1$ for all $\x_i,\x_j \in \mathcal{X}$. 

Kernel-based nonparametric regression methods usually require an appropriate bandwidth to estimate the regression function. However, bandwidth selection methods developed under the assumption of independent and identically distributed (i.i.d.) errors tend to break down in the presence of correlated errors. This problem has been widely studied, see the review paper of Opsomer et al.\cite{opsomer2001nonparametric}. When prior knowledge or parametric form of the correlation structure is available or assumed, many bandwidth selection methods have been developed, for example, difference-based time series cross-validation \cite{hall2003using}, plug-in bandwidth selection for AR(1) error process \cite{francisco2004plug}, modified generalized cross-validation method for spatial data \cite{francisco2005smoothing}, far casting cross-validation  \cite{carmack2009far} and generalied correlated cross-validation  \cite{carmack2012generalised}, etc.

There are a few methods that have been developed to bypass the estimation of the error correlation structure. Most of them are based on a bimodal kernel function to mitigate the effect of error correlation \cite[e.g.][]{kim2009using, de2011kernel}. These methods are restricted to one-dimensional, equal-distance, fixed designs. Brabanter et al \cite{de2018local} first applied the bimodal kernel to remove the error correlation structures in local polynomial regression with random design. Recently, Liu and Yang\cite{liu2024kernel} and Kong et al\cite{kong2024nonparametric} adopted the bimodal kernel for both nonparametric regression function and derivativative estimation. Although the bimodal kernel can mitigate the impact of error correlation on bandwidth selection, it is effective only for short-range dependence. 

\begin{figure}[h!]
	\centering
	\subfigure[Raw data]{\includegraphics[scale=0.3]{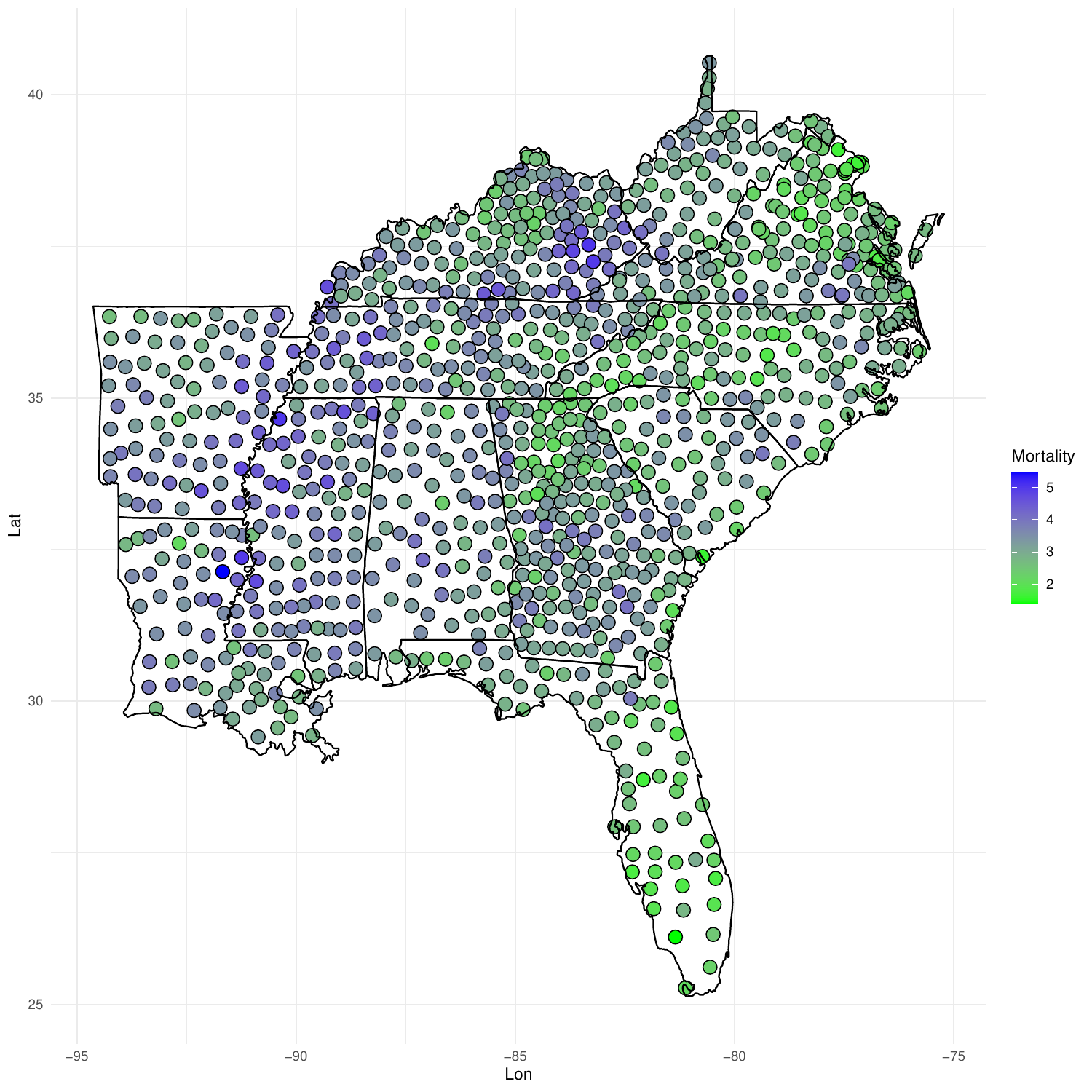}}
	\subfigure[Empirical Covariance]{\includegraphics[scale=0.3]{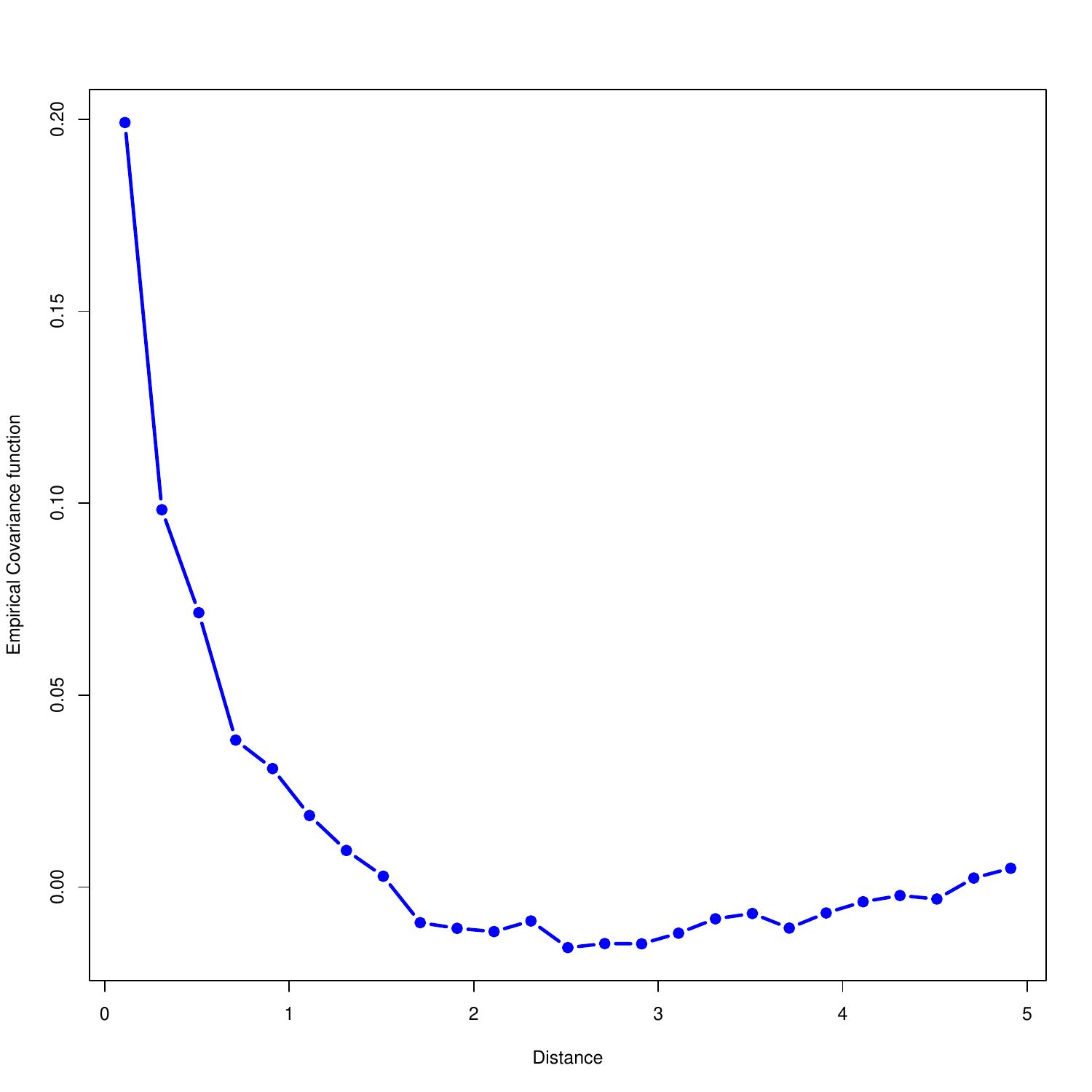}}
	\caption{The left panel is the cardiovascular disease mortality rates of 1064 counties in the southeastern United States in the year 2014. The right panel is the empirical covariance versus the distance ($\times10^2$km).}
	\label{figure_app_raw}
\end{figure}

The major motivation of our article comes from \href{https://ghdx.healthdata.org/}{the Global Health Data Exchange}, which records a variety of disease data by different counties in the United States (and also other countries). These data sets can be useful for analyzing the spatial distribution of various type of disease for the interests of the epidemiologist, including \href{https://ghdx.healthdata.org/record/ihme-data/united-states-cardiovascular-disease-mortality-rates-county-1980-2014}{the Cardiovascular Mortality Rates data}, \href{https://ghdx.healthdata.org/record/ihme-data/united-states-adult-life-expectancy-county-1987-2007}{the Life Expectancy Data},  \href{https://ghdx.healthdata.org/record/ihme-data/united-states-cancer-mortality-rates-county-1980-2014}{the Cancer Mortality Rates data} etc. All these datasets conform to model \eqref{Sec1_model}, where the two-dimensional random vector $\pmb{X}_i$ represents the latitude and longitude of each county. Figure \ref{figure_app_raw}(a) illustrates the Cardiovascular Mortality Rates in the Southeastern United States at the year 2014, displaying a clear nonlinear trend in mortality rates, which indicate that a nonparametric fit would be appropriate. In addition, we applied local linear regression with bandwidth chosen via generalized cross-validation and then used the residuals to compute the empirical error covariance. Figure \ref{figure_app_raw}(b) plots the empirical error covariance against the distances, revealing that errors may exhibit strong dependence. In Section \ref{Sec_real_data_application}, we will demonstrate that the bimodal kernel function fails to mitigate the impact of error correlation on bandwidth selection in the motivated example. Moreover, another two data examples --- the Life Expectancy and the Colon \& Rectum cancer mortality rate, will be present in the Supplementary file.

In this paper, we introduce a new class of kernel functions and propose a procedure for selecting bandwidth in kernel-based regression. We use local linear regression as an example. Under mild conditions, it is shown that the selected bandwidth is asymptotically optimal in the sense of minimizing the Mean Integrated Square Error. The proposed selection procedure performs better for more complex correlation structures and remains effective for short-range dependent errors. 

While the proposed bandwidth selection procedure does not require knowing the correlation structure, the estimation of the error covariance function has numerous potential applications, including but not limited to its use in kernel ridge regression with incorporated error covariance \cite{dang2023generalized}, tuning parameter selection in derivative estimation \cite{liu2022generalized}, and in real data applications where the correlation structure reflects relationships among various observed times/locations, such as our motivated examples. However, most nonparametric estimations of the error covariance function rely on assumptions of fixed equally spaced designs, $m$-dependence, strong mixing with exponential decay rate, auto-regression\cite{hall2003using}\cite{altman1990kernel,altman1993estimating,park2006simple,yang2019st,cui2021estimation}. In random design settings, the methods proposed by Hall et al\cite{hall1994nonparametric} and Choi et al\cite{choi2013nonparametric}, do not assume a nonparametric form of the regression function.

Our second contribution in this paper is the development of nonparametric estimations for the error covariance function in random design using kernel smoothing. We investigate the theoretical properties of the estimation without assuming any covariance structure. Our proposed methodology has the potential for both methodological and scientific applications.

The paper is structured as follows. Section \ref{Sec_band} describes basic assumptions on the error correlation. We then introduce the new class of kernel functions and demonstrate the procedure for bandwidth selection using local linear regression. The estimation of the error covariance function is introduced and studied in Section \ref{Sec_Cov_est}. 
In Section \ref{Sec_Simulation}, we evaluate the performance of our method through simulation studies. Section \ref{Sec_real_data_application} is dedicated to a real-data application. Potential directions for future research are discussed in Section \ref{Sec_discuss}. Most of the key theoretical results are included in the Appendix and all proofs, additional simulation results and applications are given in the Supplementary file.

\section{Bandwidth Selection}\label{Sec_band}

\subsection{Method and theoretical properties}
First, we introduce some notations for the rest of the paper. We will use $\Vert\cdot\Vert$ to denote the standard Euclidean norm and use notation $a_n  \asymp b_n$ to indicate that both $a_n/b_n$ and $b_n/a_n$ are bounded (in probability) as $n\rightarrow\infty$. Denote the unit vector with 1 in the first position by $\pmb{e}_1=(1,0,\ldots,0)^T$. Let $\Y=(Y_1,\ldots,Y_n)^T$ be the vector of all observations and $\X_{\x}$ be the design matrix at point $\x\in\mathcal{X}$ with $i$-th row to be $(1 ,(\X_i-\x)^T)$. 

Considering the nonparametric regression model \eqref{Sec1_model}, it is common to assume that observations located close to each other are more related than those further apart. For instance, Brabanter et al\cite{de2018local} assumes a short-range correlation with statistical dependence that decays rapidly to 
0 as the distance between points increases. In this article, we allow for a slower decay rate, which goes beyond the assumption of short-range dependence, and propose the following assumptions.

\begin{description}
	\item[A1:]  For some positive number $0<\alpha\leq 1$, there exist positive constants $C_{\rm max}$ and $C_{\rho}$ such that, 
	\begin{align*}
		{n^{\alpha}\int \vert \rho_n(\Vert\pmb{u}\Vert)\vert d\pmb{u} < C_{\rm max} \text{ and } } \lim_{n\rightarrow\infty} n^{\alpha}\int  \rho_n(\Vert\pmb{u}\Vert) d\pmb{u} =C_{\rho}. 
	\end{align*}
	\item[A2:]  There exists sequence $\tau_n = O\left(n^{-\frac{\alpha}{D+4}}\right)$, such that 
	\begin{equation*}
		\lim_{n\rightarrow\infty} n^{\alpha}\int_{\Vert \pmb{u}\Vert \geq \tau_n} \vert\rho_n(\Vert\pmb{u}\Vert) \vert d\pmb{u} = 0.
	\end{equation*}
	\item[A3:]  For all $i,j,k,l$, 
	\begin{align*}
		\cov(\varepsilon_i\varepsilon_j,\varepsilon_k\varepsilon_l\vert \X_i,\X_j,\X_k,\X_l)=& \
		\cov(\varepsilon_i,\varepsilon_k\vert \X_i,\X_k )\cov(\varepsilon_j,\varepsilon_l\vert \X_j,\X_l)\\&+\cov(\varepsilon_i,\varepsilon_l\vert \X_i,\X_l)\cov(\varepsilon_j,\varepsilon_k\vert \X_j,\X_k)\end{align*}
	
\end{description}
Assumption A3 is satisfied, for example, when the errors follow a normal distribution. 
Assumption A1 indicates that the correlation function  $\rho_n(\cdot)$ depends on $n$ and that $\int \rho_n(\Vert\pmb{u}\Vert) d\pmb{u} $ should vanish at a rate not slower than $O(n^{-\alpha})$. Assumption A2 shows that $\int \vert\rho_n(\Vert\pmb{u}\Vert) \vert d\pmb{u}$ is dominated when $\Vert \pmb{u}\Vert \leq \tau_n$. Note that when $\alpha=1$ and $\tau_n$ is of the order  $n^{-\frac{1}{D}}$, $\rho_n(\cdot)$ corresponds to the short-range correlation function in Brabanter et al\cite{de2018local}. However, assumptions A1 and A2 are much looser, which allows a slower decay rate of $\int  \rho_n(\Vert\pmb{u}\Vert) d\pmb{u} $ and $\rho_n(\Vert\pmb{u}\Vert)$ respectively.  For instance, the correlation function  
\begin{align*}
	\rho_n(\Vert \pmb{u}\Vert) &=e^{-cn^{\frac{\alpha}{D}}\Vert \pmb{u}\Vert} ,\quad c>0, 
\end{align*}
satisfies the assumptions A1 and A2. The decay rate is slower than that of the (short-range) correlation function,  $\rho_n(\Vert\pmb{u}\Vert) = e^{-cn^{\frac{1}{D}}\Vert \pmb{u}\Vert}$, of the exponential model\cite{cressie2015statistics}.  Moreover, assumption A2 allows $\tau_n$ to be at most of the order $n^{-\frac{\alpha}{D+4}}$. For example, we can construct the correlation function 
\begin{align*}
	\rho_n(\Vert\pmb{u}\Vert) &= \left(e^{-n^{2}\Vert\pmb{u}\Vert} +n^{-\frac{3\alpha}{D+4}}\Vert\pmb{u}\Vert -n^{-\frac{\alpha}{D+4}}\Vert\pmb{u}\Vert^3 \right)I\left(\Vert\pmb{u}\Vert\leq n^{-\frac{\alpha}{D+4}}\right), 
\end{align*}
satisfies assumptions A1 and A2 and has a slower decay rate of $\int  \rho_n(\Vert\pmb{u}\Vert) d\pmb{u}$.

The following bandwidth selection procedure is adapted to the situation with a slower decay rate of both $\int  \rho_n(\Vert\pmb{u}\Vert) d\pmb{u}$ and  $\rho_n(\Vert\pmb{u}\Vert)$. The method can be widely applied to  kernel-based nonparametric regression, like Nadaraya-Watson regression, Gasser-M{\"u}ller regression, local linear regression, etc. Here, we use the local linear regression as an example to show the procedure. The estimator for  the mean function $\mu$ in  the model \eqref{Sec1_model} at an arbitrary point $\x\in\mathcal{X}$ can be written as 
\begin{equation}
	\widehat{\mu}_{h,K}(\x) = \pmb{e}^T_1(\X_{\x}^T\W_{\x}\X_{\x})^{-1}\X_{\x}^T\W_{\x}\Y, \label{Sec2_local_linear}
\end{equation}
where $\W_{\x}$ is the $n\times n$ diagonal matrix of weights, i.e., $\text{diag}\{K(\Vert\X_i-\x\Vert/h)/h\}$, with kernel function $K$ and bandwidth $h$.

A natural and computationally convenient method to estimate $\widehat{\mu}_{h,K}(\x)$, is to use the Mean Integrated Square Error (MISE) criterion, given by
\begin{align*}
	\MISE(h,K) = \E\left[\int f(\x)\left(\widehat{\mu}_{h,K}(\x)-\mu(\x)\right)^2 d\x\big\vert \X\right],
\end{align*}
where $\X = \{\X_1,\ldots,\X_n\}$, $h$ is the selected bandwidth and $K$ is the kernel function used for estimating $\mu$. To minimize the MISE, we need to select a bandwidth $h$ and a kernel function $K$. It is well-known that errors are correlated, general smooth parameter selection methods such as Mallow's $C_p$\cite{mallows1973some} or Generalized Cross Validation (GCV) fail to obtain a reasonable bandwidth. 
Bimodal kernels have been proposed for bandwidth selection in univariate nonparametric regression\cite{kim2009using, de2011kernel, de2018local}. They necessitate the kernel function $K$ to satisfy $K(0)=0$ and to be Lipschitz continuous at $0$. However, when the decay rate of correlation function $\rho_n(\cdot)$ is slow, mere Lipschitz continuity at $0$ may not be sufficient for effectively eliminating the correlation structure.  

To accommodate the more complex correlation structures in multivariate nonparametric regression, we introduce a new class of kernel functions. Let $c_1<c_2$ be two positive numbers. 
We consider the kernel function $K(\cdot)$ depending on the Euclidean norm of vector $\pmb{u}=(u_1,\ldots,u_D)^T$, that satisfies the following conditions: 
\begin{description}
	\item[C1:] $K(\Vert\pmb{u}\Vert)>0$ for $c_1<\Vert\pmb{u}\Vert<c_2$ and $K(\Vert\pmb{u}\Vert)=0$ otherwise. 
	\item[C2:] $\int K(\Vert\pmb{u}\Vert)d\pmb{u}=1$ and $\int u_iu_jK(\Vert\pmb{u}\Vert)d\pmb{u}=0$ for any $i\neq j$.
	\item[C3:] $K$ is Lipschitz continuous on interval $[c_1,c_2]$.
\end{description}

We denote such kernel function by $K_z(\cdot)$ in the rest of the manuscript, where the subscript $z$ indicates that the kernel function is zero in a neighborhood of the origin and non-zero outside this neighborhood. It is worth noting that $ K_z(\cdot) $ is defined on an annular region $ c_1 < \Vert\pmb{u}\Vert < c_2 $. This feature plays a crucial role in mitigating the effects of error correlation. Denote the residual sum of squares as $\text{RSS}(h,K_z) = \frac{1}{n}\sum_{i=1}^n\left[Y_i - \widehat{\mu}_{h,K_z}(\pmb{X}_i)\right]^2 $. Suppose the local linear estimator at $x = \pmb{X}_i$ can be written as $ \mu(\pmb{X}_i) = \sum_{s=1}^{n}c_{is}Y_s $. Then, it can be shown that
\begin{equation}\label{Sec2_approximate_MISE}
	\text{MISE}(h,K) \approx \text{E}\left[\text{RSS}(h,K)\big\vert \pmb{X} \right] - \sigma^2 + \frac{2\sigma^2}{n}\sum_{i\neq s}c_{is}\rho_n\left(\Vert\pmb{X}_i-\pmb{X}_s\Vert\right)
\end{equation}
from the supplementary. In local linear regression, the linear coefficient $c_{is}$ is largely determined by $K\left(\frac{\Vert\pmb{X}_i-\pmb{X}_s\Vert}{h}\right)$. Therefore, if we use the kernel function $ K_z $, when $ \pmb{X}_i $ is close to $ \pmb{X}_s $, $ \rho_n\left(\Vert\pmb{X}_i-\pmb{X}_s\Vert\right) $ is large, but $ c_{is} $ becomes very small because $ K_z $ is zero in the neighborhood of the origin. On the other hand, when $ \pmb{X}_i $ is relatively far from $ \pmb{X}_s $, $ \rho_n\left(\Vert\pmb{X}_i-\pmb{X}_s\Vert\right) $ is small, and $ c_{is} $ remains a positive constant. As a result, if we pick a appropriate large enough $c_1$ such that $c_{is}$ is almost zero when $\rho_n\left(\Vert\pmb{X}_i-\pmb{X}_s\Vert\right)$ is large, the last term on the right hand side of above formula becomes small relative to $ \text{E}\left[\text{RSS}(h,K_z)\big\vert\pmb{X} \right] $. Therefore, intuitively, minimizing $ \text{RSS}(h,K_z) $ allows $ \text{MISE}(h,K) $ to be asymptotically minimized when the kernel function is fixed as $ K = K_z(\cdot) $. However, the kernel $K_z$ is not MISE-optimal. As a well-known result, the Epanechnikov kernel (denoted as $K_o$) can lead to the minimization of the asymptotic MISE. Therefore, we need to find the bandwidth that asymptotically minimizes $\text{MISE}(h,K_o)$. Fortunately, the factor method \cite{muller1987bandwidth} has been developed such that the ratio of two asymptotically MISE-optimal bandwidths for two different kernel functions is a constant, which depends only on the kernel functions. Thus, we can employ the factor method to obtain the asymptotically MISE-optimal bandwidth for the kernel function $K_o$.

Based on the motivation above, we propose the following procedure to select $h$ for estimating $\mu$.
\begin{description}
	\item[Step 1:] Pick a kernel $K_z$ with a appropriate large enough $c_1$ value, select a bandwidth, denoted as $\widehat{h}(K_z)$, that minimizes the residual sum of squares:
	\begin{equation}\label{Sec2_RSS}
		\RSS(h,K_z) = \frac{1}{n}\sum_{i=1}^n\left(Y_i - \widehat{\mu}_{h,K_z}(\X_i)\right)^2,
	\end{equation}
	\item[Step 2:] Calculate
	\begin{equation} \label{Sec2_h_factor}
		\widehat{h}(K_o) = \widehat{h}(K_z)\left(\frac{\mu(K_o^2)\mu_2(K_z)^2}{\mu_2(K_o)^2\mu(K_z^2)}\right)^{\frac{1}{D+4}},
	\end{equation}
	where $\mu_2(K)=\int u_1^2K(\Vert\pmb{u}\Vert)d\pmb{u}$ and $\mu(K^2)=\int K^2(\Vert\pmb{u}\Vert)d\pmb{u}$.
	\item[Step 3:] Substitute $K=K_o$ and $h=\widehat{h}(K_o)$ into \eqref{Sec2_local_linear} to obtain the estimate of local linear regression function.
\end{description}

Note that the above procedure does not require estimating correlation function $\rho_n(\cdot)$. 
This results in a flexible method for estimating the regression function in the presence of correlated errors. In step 1, we apply the kernel function $K_z$ to eliminate correlation, the selected $\widehat{h}(K_z)$ can asymptotically minimize MISE when the kernel function is $K_z$. However, the kernel $K_z$ is not MISE-optimal with fixed bandwidth, in step 2, we employ the factor method to determine the bandwidth suitable for the MISE-optimal kernel function $K_o$. 

The following theorem presents the asymptotic results regarding the selected bandwidth $\widehat{h}(K_o)$ and the MISE achieved.

\begin{theorem}\label{Sec2_Thm}
	Suppose assumptions A1 to A3 hold. Minimizing \eqref{Sec2_RSS} with the candidate set of bandwidth $h \in \mathcal{H}_n= (a_1n^{-\frac{\alpha}{D+4}},a_2n^{-\frac{\alpha}{D+4}})$, where $a_1$ and $a_2$ are some positive constant, we obtain $\widehat{h}(K_o)$ via \eqref{Sec2_h_factor}. Let $h^*$ be the bandwidth that minimizes MISE with the kernel function $K_o(\cdot)$. Then,
	\begin{equation*}
		\frac{\MISE\left(\widehat{h}(K_o),K_o\right)}{\MISE\left(h^*,K_o\right)} = 1+o_p(1), \quad
		\frac{\left\vert\widehat{h}(K_o)-h^*\right\vert }{h^*} = o_p(1)
	\end{equation*}
	and 
	\begin{equation*}
		\MISE\left(\widehat{h}(K_o),K_o\right) = C_1n^{-\frac{4\alpha}{D+4}} \left[\mu(K_o^2)^2\mu_2(K_o)^D\right]^{\frac{2}{D+4}}+ o_p\left(n^{-\frac{4\alpha}{D+4}}\right),
	\end{equation*}
	where 
	\begin{equation}
		C_1 =\left\{
		\begin{aligned}
			&2\left(\frac{\Delta_f}{2}\right)^{\frac{D}{D+4}}\left(\sigma^2C_\rho\right)^{\frac{4}{D+4}}, \quad  0<\alpha<1,\\
			&2\left(\frac{\Delta_f}{2}\right)^{\frac{D}{D+4}}\left(\sigma^2(C_\rho+m(\mathcal{X}))\right)^{\frac{4}{D+4}}, \quad  \alpha=1,
		\end{aligned}
		\right. \nonumber
	\end{equation}
	with $m(\mathcal{X}) = \int_{\x\in\mathcal{X}} 1d\x$ and $\Delta_f = \int f(\x)\left(\sum_{d=1}^{D}\frac{\partial^2 \mu(\x)}{\partial x_d^2}\right)d\x$.
\end{theorem}
Theorem \ref{Sec2_Thm} states that through the aforementioned bandwidth selection procedure, we can achieve an optimal estimation of $\mu(\boldsymbol{\cdot})$ in terms of minimizing the asymptotic MISE. Building upon the proof of the above theorem, the following result emerges.
\begin{corollary}
	Suppose that conditions in Theorem \ref{Sec2_Thm} hold. Let $\widehat{\mu}_{h,K_o}(\cdot)$ be the estimator \eqref{Sec2_local_linear} with the bandwidth $\widehat{h}(K_o)$ and kernel function $K_o(\cdot)$. Then
	\begin{equation*}
		\widehat{\mu}_{h,K_o}(\x) - \mu(\x) = O_p(n^{-\frac{2\alpha}{D+4}}) \quad \text{  for  any  } \x\in \mathcal{X}.
	\end{equation*} 
\end{corollary}

\subsection{Practical issues.}

One of the practical questions is how to determine the function $K_z$. A simple option is to employ a fourth-degree polynomial function:
\begin{equation}\label{Sec2_kernel}
	K_z(\Vert\pmb{u}\Vert) = (A_z\Vert\pmb{u}\Vert^3 + B_z\Vert\pmb{u}\Vert^2 + C_z\Vert\pmb{u}\Vert + D_z)I(c_1\leq\Vert\pmb{u}\Vert\leq c_2),
\end{equation}
where $A_z$, $B_z$, $C_z$, and $D_z$ are constants. Assuming $c_1$ and $c_2$ are fixed, to determine the coefficients $A_z$, $B_z$, $C_z$, and $D_z$ involves minimizing two quantities. There are two possible choices. One of the choices is to minimize $\mu(K_z^2)$ to control the variance of MISE for the kernel $K_z$. The other is to minimize $\left[\mu(K_z^2)^2\mu_2(K_z)^D\right]$, which leads to the asymptotic minimization of $\MISE(h,K_z)$. %We have provide R code to obtain $A,B,C,D$ for both ways. 

Secondly, the selection of $c_1$ would be important in real data analysis. From our simulation, we find that $c_2=c_1+1/2$ works pretty well for various situations. In addition, we find that if $c_1$ is well selected, then the gap between $c_1$ and $c_2$ would be much less concern in the simulation. Therefore, it is recommend to use $c_2 = c_1 + 1/2$ in practice. The choice of $c_1$ plays a important role for selecting the bandwidth $h$. To select bandwith, we will propose a ``elbow method" for the purpose. 

From the Corollary A.2 in appendix, we showed that the bandwidth that can minimize the leading term of $\text{MISE}(h,K)$ should be proportional to $\left(\mu(K^2)\big/\mu_2(K)^2\right)^{\frac{1}{D+4}}$, therefore, we should have 
\begin{equation*}
	\bar{C} =  \left(\frac{\mu(K_z^2)}{\mu_2(K_z)^2}\right)^{\frac{1}{D+4}}\bigg /\widehat{h}(K_z),
\end{equation*}
where $\bar{C}$ is a constant only depends on $\rho_n$, $\mu$, and $f$. Suppose the kernel function $K_z$ is applied, and $c_1$ is large enough such that the last term of \eqref{Sec2_approximate_MISE} is negligible, then intuitively, the bandwidth $\widehat{h}(K_z)$ should approximately minimizes $\text{MISE}(h,K_z)$ as well. Therefore, at this time, the ratio of $\widehat{h}(K_z)$ and $\left(\mu(K_z^2)\big/\mu_2(K_z)^2\right)^{\frac{1}{D+4}}$ should be approximately be a constant, or at least somehow stable.  

Based on the above motivation and our large empirical studies, we propose a empirical visual method, called ``elbow method", for roughly pick $c_1$ values. 
\begin{itemize}
	\item[1.] Put a candidate set with $N_1$ of $c_1$ values as $\mathcal{C}_1=\{c^{(1)}_1,c^{(2)}_1,...,c^{(n_1)}_1..,c^{(N_1)}_1\}$, where $N_1$ is a positive integer and $\mathcal{C}_2=\mathcal{C}_1+0.5$.
	\item[2.] For combination of $(c^{(n_1)}_1,c^{(n_1)}_2)$ values, we compute the corresponding kernel function $K_z$ as (3.6) such that the condition C1 to C3 is satisfied and $\left[\mu(K_z^2)^2\mu_2(K_z)^D\right]$ is minimized.
	\item[3.] For each $K_z$ kernel, we use \eqref{Sec2_kernel} to find the $\widetilde{h}(K_z)$, and compute the quantity $\left(\frac{\mu(K_z^2)}{\mu_2(K_z)^2}\right)^{\frac{1}{D+4}}$ as well.
	\item[4.] Compute the ratio $\left(\frac{\mu(K_z^2)}{\mu_2(K_z)^2}\right)^{\frac{1}{D+4}}\bigg /\widehat{h}(K_z)$ for each $K_z$ kernel and plot it against $\{c^{(1)}_1,c^{(2)}_1,...,c^{(n_1)}_1..,c^{(N_1)}_1\}$.
	\item[5.] Pick the \textbf{first} ``elbow area" such that the ratio is roughly stablized for the first several $c_1$ values as the choice to find the kernel $K_z$ for removing the effect of $\rho_n$ on bandwidth selection.
\end{itemize}

\begin{figure}[h!]
	\centering 
	\subfigure{\includegraphics[scale=0.5]{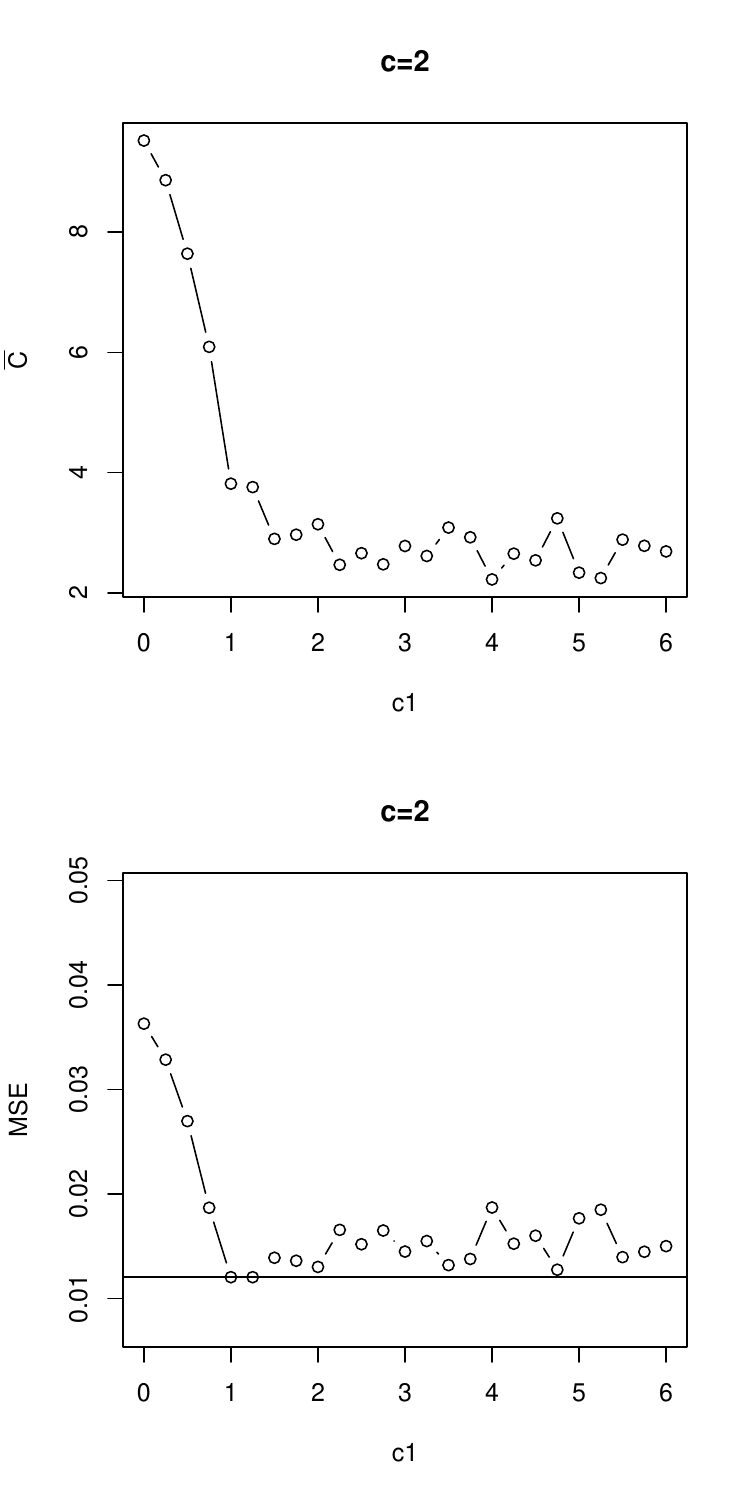}}\qquad 
	\subfigure{\includegraphics[scale=0.5]{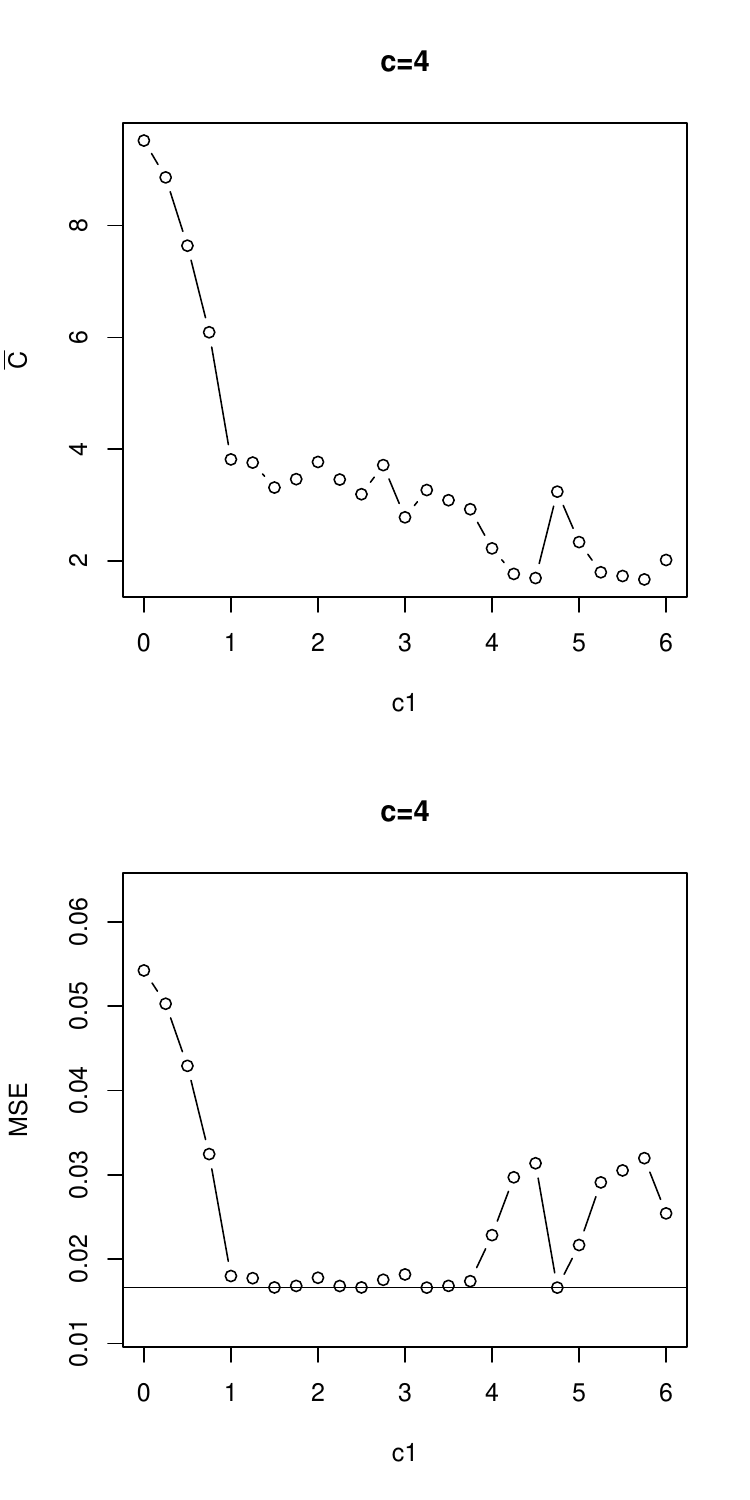}}\\
	\caption{The plots comes from a typical example of our simulation trials ($seed=100$) with spherical model and $D=2$. In the two bottom plots, the dark solid line is the minimum MSE value we can achieve.   }
	\label{figure_plot_c1}
\end{figure}

Figure \ref{figure_plot_c1} illustrates how to determine $c_1$. The top two plots display the values of $ \bar{C} $ for different choices of $ c_1 $ under a spherical correlation model with $D=2$ and $c=2$ or $c=4$. The bottom two plots show the corresponding MSE values for each $c_1$. By comparing the top-left and bottom-left plots or the top-right and bottom-right plots, it is evident that the "elbow region" of $c_1$ values yields an optimal MSE.

%%%%%%%%%%%%%%%%%%%%%%%%%%%%%%%%%%%%%%%%%%%%%%%%%%%%%%%%%%%%%%%%%%%%%%%%%%%%%%%%%%%%%%%%%%%%%%%%%%%%%%%%%%%%%%%%%%%%%%%%%%%%

\section{Estimation of the Covariance Function}\label{Sec_Cov_est}

In this section, we propose employing a nonparametric smoothing method to estimate the error covariance function $C_n(t)=\sigma^2\rho_n(t)$ under assumptions A1 to A3. As $\rho_n(\cdot)$ varies with $n$, we further need the following assumption.
\begin{description}
	\item[A4:] $\rho_n(t) = \rho(n^{\frac{\alpha}{D}}t)$ for any $t\geq 0$, where $\rho(\cdot)$ is a valid correlation function unrelated to $n$, with bounded first and second derivatives.
\end{description}

Some common correlation functions satisfy assumption A4. For instance,
\begin{align*}
	\text{Spherical:} \qquad \rho_n(t) &= \left(1-\frac{3n^{\frac{\alpha}{D}}t}{2c} + \frac{n^{\frac{3\alpha}{D}}t^3}{2c^3}\right) I(t\leq cn^{-\frac{\alpha}{D}})\\
	\text{Exponential:}\qquad  \rho_n(t) &= e^{-cn^{\frac{\alpha}{D}}t}\\
	\text{Inverse quadratic:}\qquad \rho_n(t) &= \frac{1}{1+cn^{\frac{2\alpha}{D}}t^2},
\end{align*} 
where $c$ is some positive number. 

Suppose the assumptions A1 to A4 hold, we nonparametrically estimate covariance function $C_n(\cdot)$ use 
\begin{equation*}
	\widehat{C}_n(t) = \frac{ \sum_{i, j=1}^{n}\widehat{\varepsilon}_i\widehat{\varepsilon}_jK_C\left(\frac{t-\Vert \X_i-\X_j\Vert}{b}\right)}{ \sum_{i, j=1}^{n}K_C\left(\frac{t-\Vert \X_i-\X_j\Vert}{b}\right)},
\end{equation*}
where $\widehat{\varepsilon}_{i}=Y_i - \widehat{\mu}_{h,K_o}(\X_i)$, $\widehat{\mu}_{h,K_o}(\X_i)$ is defined as \eqref{Sec2_local_linear} with kernel function $K_o$, $b$ is the bandwidth to control the smoothness of $\widehat{C}_n(t)$, and  $K_{C}$ is kernel function satisfying
\begin{equation*}
	\begin{aligned}
		&K_C(u)>0\text{ for }u\in(-1,1)\text{ and }K_C(u)=0\text{ otherwise},\\
		& \int K_C(u)du = 1,\ \int uK_C(u)du =0,\ \text{ and }\int u^DK_C(u)du  > 0.
	\end{aligned}
\end{equation*}

The estimator $\widehat{C}_n(t)$ is closely related to the estimate of regression function in model \eqref{Sec1_model} of Hall et al\cite{hall1994nonparametric}. However, Hall et al\cite{hall1994nonparametric} assume one-dimensional increasing domain asymptotics which does not fit with the assumptions of model \eqref{Sec1_model}, whereas we assume infill asymptotics with a finite domain of $\X_i$, leading to a natural difference in theoretical investigation. To analyze the estimator $\widehat{C}_n(t)$, we introduce the following estimator:
\begin{equation*}
	\widetilde{C}_n(t) = \frac{ \sum_{i, j=1}^{n}\varepsilon_i\varepsilon_jK_C\left(\frac{t-\Vert \X_i-\X_j\Vert}{b}\right)}{ \sum_{i, j=1}^{n}K_C\left(\frac{t-\Vert \X_i-\X_j\Vert}{b}\right)},
\end{equation*} 
which assumes $\varepsilon_i$'s are observable. 

\begin{theorem}\label{Sec3_Thm3.1}
	Suppose assumptions A1 to A4 hold and bandwidth $b$ satisfies $bn^{\frac{\alpha}{D}}\rightarrow 0$ and $bn^{\alpha(1+1/D)}\rightarrow \infty$. Then, if $t \asymp n^{-\frac{\alpha}{D}}$,
	\begin{align*}
		\mbox{\rm Bias}\left(\widetilde{C}_n(t)\big\vert \X\right)&=\E\left(\widetilde{C}_n(t)\big\vert \X\right)  - C_n(t) = O_p(b^2n^{\frac{2\alpha}{D}}), \\
		\var\left(\widetilde{C}_n(t)\big\vert \boldsymbol{X}\right) & = O_p\left(\frac{1}{bn^{\alpha+\alpha/D}}\right).
	\end{align*}
\end{theorem}
Theorem \ref{Sec3_Thm3.1} shows that if we control $b$ such that $b=o(n^{-\frac{\alpha}{D}})$ and $bn^{\alpha(1+1/D)}\rightarrow \infty$, then $\widetilde{C}_n(t)$ will be a consistent estimator of $C_n(t)$ for any $t \asymp n^{-\frac{\alpha}{D}}$. 

We must pay careful attention to the one-dimensional case with short-range correlation. If we restrict $\alpha=1$ and $D=1$, then clearly, $b$ needs to be $o(n^{-1})$. On the other hand, since $X_i$ are random samples from a bounded $f$, it can be argued that $\underset{1\leq i\leq n}{\min}\left\vert X_i-X_j\right\vert  \asymp n^{-1}$. Therefore, if $b=o(n^{-1})$ and $t \asymp n^{-1}$, it can be shown that 
\begin{equation*}
	\E\left(\sum_{i\ne j}^nI\left(\left\vert X_i-X_j\right\vert \in (t-b,t+b) \right)\bigg\vert \X\right) = o_p(1).
\end{equation*}
It means that as $n\rightarrow\infty$, the expected number of distances $\vert X_i-X_j\vert$ inside the interval $[t-b,t+b]$ goes to 0. In that case, we are not able to obtain a valid $b$ for the consistent estimator of $\rho_n(t)$. 

Next, we state the difference between $\widehat{C}_n(t)$ and $\widetilde{C}_n(t)$ in the following results.
\begin{theorem}\label{Sec3_Thm3.2}
	Suppose that conditions in Theorem \ref{Sec3_Thm3.1} are satisfied. We obtain $\widehat{\varepsilon}_i$ with bandwidth $h$ and kernel function $K_o$. Then, 
	\begin{equation*}
		\widehat{C}_n(t) - \widetilde{C}_n(t) = O_p\left(\max\left\{h^4,\frac{1}{n^\alpha h^D}\right\}\right). 
	\end{equation*}
\end{theorem}
\begin{corollary}\label{Sec3_Coro3.1}
	Suppose that conditions in Theorem \ref{Sec3_Thm3.1} are satisfied. We obtain $\widehat{\varepsilon}_i$ with bandwidth $\widehat{h}(K_o)$ and kernel function $K_o$. Then, if $t \asymp n^{-\frac{\alpha}{D}}$, 
	\begin{equation*}
		\widehat{C}_n(t) - C_n(t) = O_p(n^{-\frac{4\alpha}{D+4}}) + O_p(b^2n^{\frac{2\alpha}{D}}) + O_p\left(\frac{1}{\sqrt{bn^{\alpha+\alpha/D}}}\right).
	\end{equation*}
\end{corollary}

From Corollary \ref{Sec3_Coro3.1}, it can be shown that choice of $b$ matters in practice for the accuracy of $\widehat{C}(t)$. Next, we define two different variance estimators which will serve for the purposes of choosing $b$ later on:
\begin{align}
	\widehat{\sigma^2} &= \frac{1}{n}\sum_{i=1}^n\left(Y_i - \widehat{\mu}_{h,K_o}(\X_i)\right)^2 \label{Sec3_VarEst1}, \quad \text{and}\\
	\widetilde{\sigma^2} &= \widehat{C}_n(0) = \frac{ \sum_{i, j=1}^{n}\left(Y_i - \widehat{\mu}_{h,K_o}(\X_i)\right)\left(Y_j - \widehat{\mu}_{h,K_o}(\X_j)\right)K_C\left(\frac{\Vert \X_i-\X_j\Vert}{b}\right)}{ \sum_{i,j =1}^{n}K_C\left(\frac{\Vert \X_i-\X_j\Vert}{b}\right)}.\label{Sec3_VarEst2}
\end{align}
\begin{theorem}\label{Sec3_Thm3.3}
	Suppose that assumptions A1 to A4 hold.  We obtain $\widehat{\mu}_{h,K_o}(\X_i)$ with some bandwidth $h$ and kernel $K_o$, then the variance estimators \eqref{Sec3_VarEst1} and \eqref{Sec3_VarEst2} satisfy 
	\begin{align}
		\widehat{\sigma^2}  &= \sigma^2 + O_p\left(h^4+\frac{1}{\sqrt{n^\alpha h^D}}\right) \label{Sec3_Var1_property}, \quad \text{and}\\
		\widetilde{\sigma^2} &= \sigma^2 + O_p\left(h^4+\frac{1}{n^\alpha h^D}\right) +O_p(bn^{\frac{\alpha}{D}}) + O_p\left(\frac{1}{b^{D/2}n^{\alpha}} \right),\label{Sec3_Var2_property}
	\end{align}
	respectively.
\end{theorem}	

Theorem \ref{Sec3_Thm3.3} states that $\widehat{\sigma^2}$ is a consistent estimator of $\sigma^2$. Furthermore, it suggests that theoretically, if we specify $b$ such that $bn^{\frac{\alpha}{D}}\rightarrow 0$ and $bn^{{\alpha}{(1+1/D)}}\rightarrow \infty$, then $\vert\widetilde{\sigma^2}-\widehat{\sigma^2}\vert=o_p(1)$. Consequently, $\vert\widetilde{\sigma^2}-\sigma^2\vert=o_p(1)$. Moreover, if $b$ satisfies these two conditions, it also fulfills the requirements of $b$ in Corollary \ref{Sec3_Coro3.1}. Therefore, if we specify $b$ to ensure $\vert\widetilde{\sigma^2}-\widehat{\sigma^2}\vert=o_p(1)$, then both $\widetilde{\sigma^2}$ and $\widehat{C}_n(t)$ will be consistent estimators.

From equation \eqref{Sec3_Var2_property}, it is evident that a bandwidth of $h \asymp n^{-\frac{\alpha}{D+4}}$ is preferable for deriving $\widetilde{\sigma^2}$. Consequently, we employ the bandwidth $h=\widehat{h}(K_o)$ in equation \eqref{Sec2_h_factor} to calculate $\widetilde{\sigma^2}$. However, according to equation \eqref{Sec3_Var1_property}, a bandwidth of $h \asymp n^{-\frac{\alpha}{D+8}}$ is favored for estimating $\widehat{\sigma^2}$. Drawing from our extensive simulation study, we propose utilizing another bandwidth $h^T = \widehat{h}(K_o){n^{-\frac{1}{D+8}}}\big/{n^{-\frac{1}{D+4}}}$. Clearly, when $\alpha=1$, $h^T \asymp n^{-\frac{1}{D+8}}$.

Additionally, according to Theorem \ref{Sec3_Thm3.1}, if $b$ is excessively small, the bias of $\widehat{C}_n(t)$ will be minimal, yet this results in a high variation in the fitted covariance function. Conversely, with increasing in $b$, the variance of $\widehat{C}_n(t)$ decreases, but its bias begins to increase. 

Given the aforementioned analysis, we introduce a practical approach for determining $b$, which we refer to as the ``Variance-calibration" procedure:
\begin{description}
	\item[Step 1.] Specify a candidate set $\mathcal{B}=\{b_1,b_2,\ldots,b_M\}$ with $b_1<b_2<\cdots<b_M$, ensuring that the initial segment of $\mathcal{B}$ is sufficiently small to observe distinct undersmoothing in the fitted covariance function.
	\item[Step 2.] Estimate $\widehat{\sigma^2}$ using equation \eqref{Sec3_VarEst1} with bandwidth $h^T = \widehat{h}(K_o){n^{-\frac{1}{D+8}}}\big/{n^{-\frac{1}{D+4}}}$.
	\item[Step 3.] Estimate $\widetilde{\sigma^2}$ using equation \eqref{Sec3_VarEst2} with bandwidth $\widehat{h}(K_o)$.
	\item[Step 4.] Fit $C_n(t)$ using $b_1, b_2, \ldots,  b_M$, and choose the largest bandwidth such that $\left\vert\widehat{\sigma^2}-\widetilde{\sigma^2}\right\vert$ is less than a small positive value $\delta_n$. 
\end{description}
The fundamental idea is to find the largest $b$ value that still allows for a consistent estimator $\widetilde{\sigma^2}$ for $\sigma^2$ and a small bias of $\widehat{C}_n(t)$ when $t$ is close to 0. With this $b$, we can achieve a smooth fit of $C_n(t)$ in the neighborhood of 0 and the consistency of $\widetilde{\sigma^2}$ and $\widehat{C}_n(t)$. In real data applications, we also recommend combining the above procedure with visualization. That is, we plot the fitted $\widehat{C}_n(t)$ versus $t$ with different bandwidth $b$, which can assist us to observe the clear pattern of underestimation and overestimation with varied $b$'s and identify the $b$ that almost aligns $\widehat{\sigma^2}$ and $\widetilde{\sigma^2}$ and gives the smoothest curve.

\begin{figure}[h!]
	\centering
	\subfigure{\includegraphics[scale=0.5]{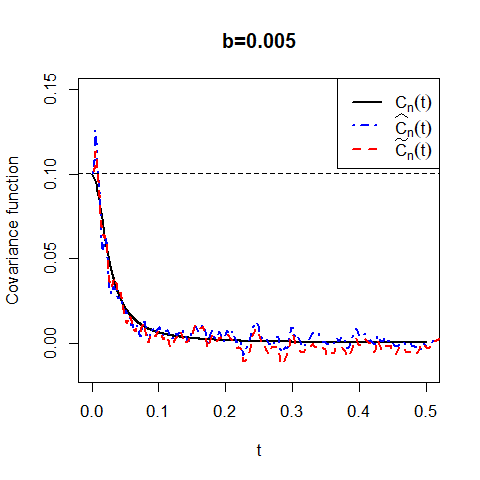}}\qquad 
	\subfigure{\includegraphics[scale=0.5]{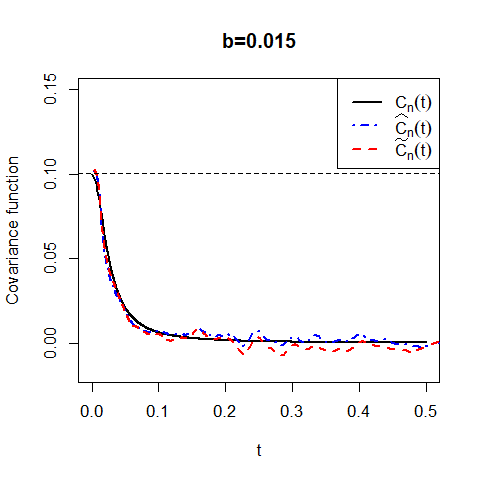}}\\
	\subfigure{\includegraphics[scale=0.5]{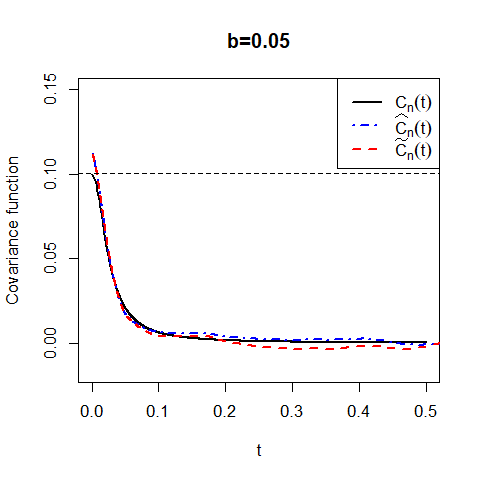}}\qquad 
	\subfigure{\includegraphics[scale=0.5]{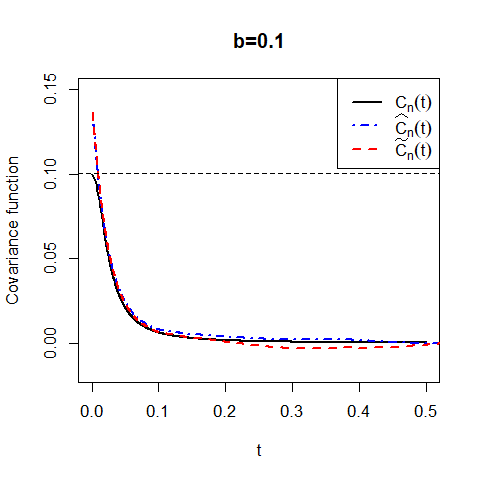}}
	\caption{Results with a two-dimensional spherical covariance with $\sigma^2=0.1$. The solid black line, blue dot-dashed line, and red dashed line represent the true covariance function $C_n(t)$, the estimated covariance functions $\widehat{C}_n(t)$ and $\widetilde{C}_n(t)$ respectively. The black dashed horizontal line stands for $\widehat{\sigma^2}$. }
	\label{figure_choose_b}
\end{figure}

Figure \ref{figure_choose_b} illustrates this procedure. The top-left panel shows the significant variance in $\widehat{C}_n(t)$ when $b$ is too small. The bottom-right panel indicates that when $b$ is too large, $\widehat{C}_n(t)$ overfit with significant bias in the neighborhood of 0. In contrast, the top-right panel shows that $b=0.015$ is just right for fitting $C_n(t)$ in the neighborhood of 0, where we have calibrated the $\widehat{\sigma^2}$ with $\widetilde{\sigma^2}$. The ``Variance-calibration" method essentially involves gradually increasing $b$ until $\widetilde{\sigma^2}$ aligns with the $\widehat{\sigma^2}$. Our simulations indicate that as long as $\widehat{\sigma^2}$ is accurately estimated, $C_n(t)$ can be well estimated in the vicinity of 0.

Theoretically, we have demonstrated that $C_n(t)$ will be consistent for $t  \asymp n^{-\frac{\alpha}{D}}$ provided that $b$ is chosen appropriately. When $n^{-\frac{\alpha}{D}} = o(t)$, $C_n(t)$ tends to zero as $n \rightarrow \infty$. Consequently, in practical applications, we recommend setting $\widehat{C}_n(t)$ to zero when $t$ exceeds a certain threshold. For instance, as depicted in Figure \ref{figure_choose_b}, our interest in covariance estimation would typically be limited to when $t \leq 0.1$, the region where $C_n(t)$ is very close to zero generally does not significantly impact data analysis. 

After choosing $b$, we take
\begin{equation}
	\widehat{\rho}_n(t) = \frac{\widehat{C}_n(t)}{\widehat{C}_n(0)}  \quad \text{   or   } \quad \widehat{\rho}_n(t) = \frac{\widehat{C}_n(t)}{\widehat{\sigma^2}}  \label{Sec3_rho_est} 
\end{equation}
as the estimation of $\rho_n(t)$. Asymptotically, the difference of $\widehat{C}_n(0)$ and $\widehat{\sigma^2}$ is negligible. Theorem \ref{Sec3_Thm3.3} also shows that as $\alpha$ decreases from $1$, the estimation of $\sigma^2$ will be less efficient. The ``Variance-calibration" uses this as a reference, which leads to a less efficient estimator of $C_n$ if $\sigma^2$ is not well-estimated. Nevertheless, the estimator $\widehat{\rho}_n$ will be less affected since the division cancel out the $\widehat{\sigma^2}$. This may be helpful for analysis of the correlation among different samples.  

One of the practical issues is that there is $O(n^2)$ of $C_n\left(\Vert\X_i-\X_j\Vert\right)$ need to be estimated, with moderately large $n$, this could be computational very time-consuming. Therefore, we suggest to estimate $C_n(t)$ at discrete points $t_0,t_1,\ldots,t_{n^*}$, where $n^*\ll n^2$ with $t_0=0$. Next, we interpolate the estimates $\widehat{C}_n(t_1),\cdots,\widehat{C}_n(t_{n^*})$ to get $\widehat{C}_n(t)$. 

Although theoretically, the bandwidth $b$ less than $t$ is desired for the consistency of $\widetilde{\rho}_n(t)$, in practice, estimation of $\rho_n(t)$ close to 0 is important since $\rho_n(t)$ is generally large when data points are close. We suggest using the boundary kernel function when the estimation location $t$ is smaller than $b$ in real data applications. From our simulation study, it is recommended to use the boundary kernel function proposed in M{\"u}ller and Wang\cite{muller1994hazard} as 
\begin{equation}
	K(t) =  \frac{12(t+1)}{(1+q)^4}\left( t(1-2q) + \frac{(3q^2-2q+1)}{2} \right)  I(\vert t\vert\leq 1) , \label{Sec3_boundary_kernel}
\end{equation}
where $q=\frac{t}{b}$ if $t<b$ and $q=1$ if $t \geq b$. If $q=1$, $K(t)$ is the Epaninikov kernel function.

\section{Numerical study}\label{Sec_Simulation}
\begin{figure}[h!]
	\centering
	\subfigure[$c=2.0$]{\includegraphics[scale=0.3]{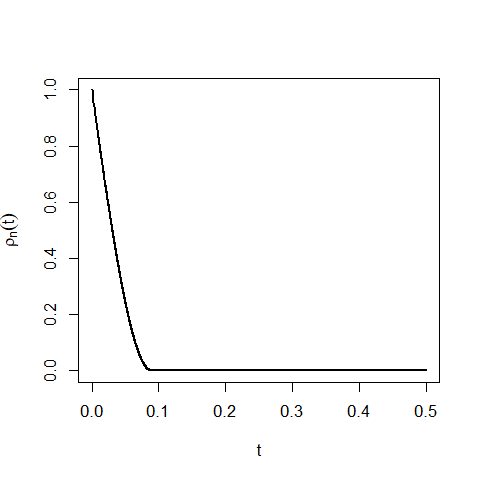}}
	\subfigure[$c=3.0$]{\includegraphics[scale=0.3]{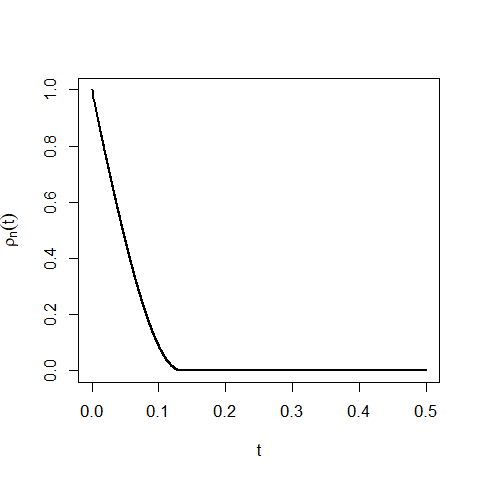}}
	\subfigure[$c=4.0$]{\includegraphics[scale=0.3]{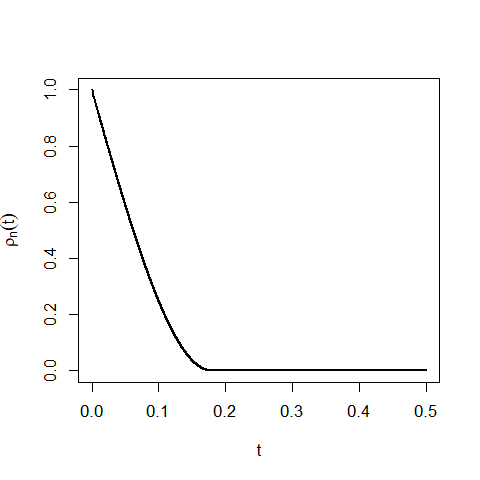}}\\
	\textit{\footnotesize Spherical model: correlation function with $c=2,3,4$, $D=2$ and $n=500$.}\\
	\subfigure[$c=2.0$]{\includegraphics[scale=0.3]{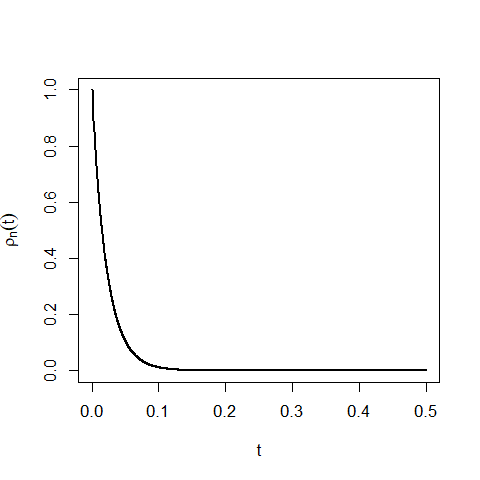}}
	\subfigure[$c=1.5$]{\includegraphics[scale=0.3]{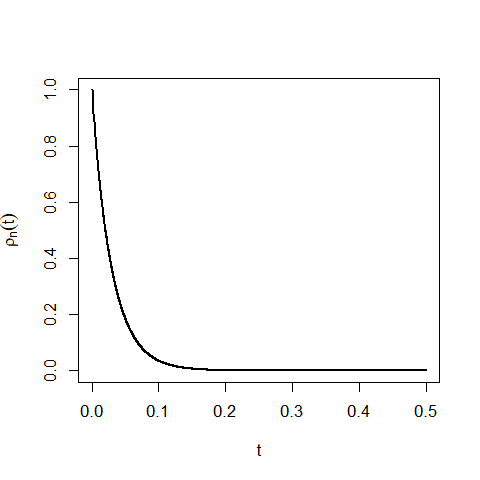}}
	\subfigure[$c=1.0$]{\includegraphics[scale=0.3]{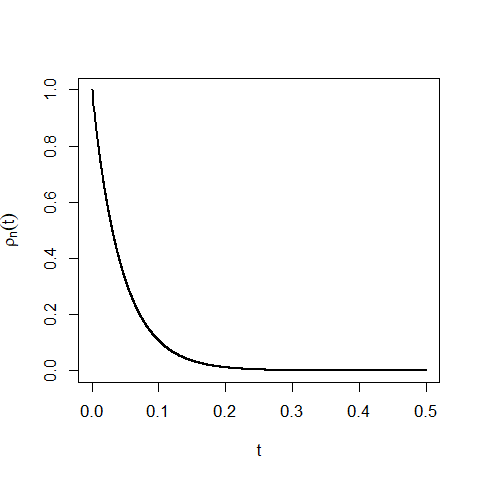}}\\
	\textit{\footnotesize Exponential model: correlation function with $c=1,1.5,2.0$, $D=2$ and $n=500$.}\\
	\subfigure[$c=7.0$]{\includegraphics[scale=0.3]{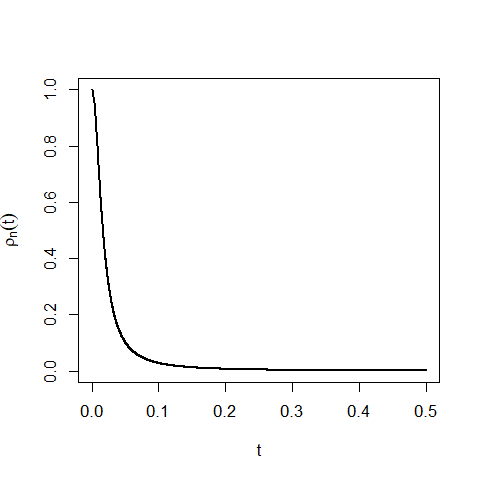}}
	\subfigure[$c=3.0$]{\includegraphics[scale=0.3]{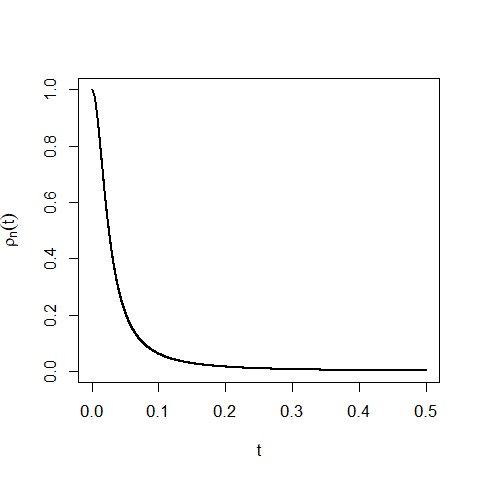}}
	\subfigure[$c=1.0$]{\includegraphics[scale=0.3]{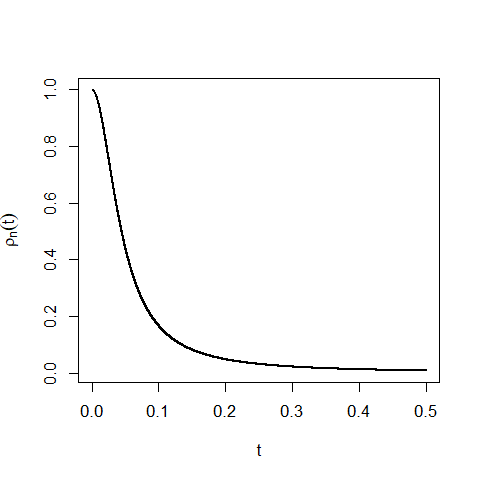}}\\
	\textit{\footnotesize Inverse quadratic model: correlation function with $c=1,3,7$, $D=2$ and $n=500$.}
	\caption{Examples of three different correlation functions with different $c$ values.}
	\label{figure_inv_quadratic_cor}
\end{figure}

In this section, we conduct simulations to evaluate the proposed method outlined in Sections \ref{Sec_band} and \ref{Sec_Cov_est}. We examine three distinct correlation functions as detailed in Section \ref{Sec_Cov_est} within our simulation study. We generate data from model \eqref{Sec1_model} considering either $D=2$ or $D=3$. The corresponding regression functions are $\mu(\mathbf{X}_i) = 2X_{i1}^2 + 2\cos(\pi X_{i2})$ for $i \in \{1, \ldots, 500\}$ and $\mu(\mathbf{X}_i) = X_{i1} + \sin(\pi X_{i2}) + 2X_{i3}^2$ for $i \in \{1, \ldots, 600\}$, where $\mathbf{X}_i$ are random vectors with each component independently generated from $\text{Uniform}(0,1)$. In each simulation scenario, we set $\sigma^2=0.1$ and $\alpha=1$. The parameter $c$ is varied to control the intensity of the error correlation. Several correlation functions with different values of $c$ when $D=2$ are illustrated in Figure \ref{figure_inv_quadratic_cor}. Due to space limitations, we only present simulation results for the case of $D=2$. Simulation results for $D=3$ will be included in the Supplementary file. 

\subsection{Estimation of regression function}\label{Sec_Simu1}
We use local linear fitting to obtain estimates of the regression function as given by \eqref{Sec2_local_linear}. The performance of the regression estimator is evaluated using the following metric:
\begin{equation*}
	\mbox{\rm MSE}_{\rm prac}(h) = \frac{1}{n}\sum_{i=1}^{n}\left( \widehat{\mu}_{h, K}(\mathbf{X}_i) - \mu(\mathbf{X}_i) \right)^2,
\end{equation*}
where $h$ represents the bandwidth used for regression estimation and $K$ is the kernel function used to derive the local linear estimator. In our simulation, we compare three different methods for bandwidth selection: (a) The Epsilon-class method proposed in Brabanter et al\cite{de2011kernel} with $\varepsilon=0.1$ (denoted as ``Eps"). (b) The Exponential kernel function proposed in the discussion of Brabanter et al\cite{de2018local} with the factor method (denoted as ``Exp"). (c) Our procedure with proposed kernel function $K_z$ with $(c_1,c_2)=(1.0,1.5)$, $(c_1,c_2)=(2.0,2.5)$, or $(c_1,c_2)=(3.0,3.5)$ (denoted as ``ZA($c_1,c_2$)").  For each pair of $c_1$ and $c_2$, we minimize $\left[\mu(K_z^2)\mu_2(K_z)\right]$ to obtain the coefficients of $K_z$. For ``Exp" and ``ZA($c_1,c_2$)", which require an optimal $K_o$ for the factor procedure, we use the product Epanechnikov kernel function as $K_o(\mathbf{u})=\prod_{d=1}^{D}K(u_d)$, with $K(u)=\frac{3}{4}(1-u^2)I(\lvert u\rvert\leq 1)$, and the same bandwidth for each direction.

We present the mean and standard deviation of $\mbox{\rm MSE}_{\rm prac}$ with the bandwidth selected from the above methods. Additionally, the mean and standard deviation of the minimum $MSE_{prac}$ that can be achieved by the product Epanechnikov kernel is computed as the reference line of each method (denoted as ``minEpan"). For each simulation scenario, we repeat 100 trials. Table \ref{MES_D_equal2} presents the mean and standard deviation of $\mbox{\rm MSE}_{\rm prac}$ for different values of $c$. ``SP", ``EXP" and ``INVQ" stand for Spherical model, Exponential model and Inverse quadratic model respectively.

\begin{table}[h!] \footnotesize
	\centering
	\caption{Mean and standard deviation (in parentheses) of $\mbox{\rm MSE}_{\rm prac}$ values ($\times 10^{-2}$) in each simulation scenario from 100 trials. The regression function is $\mu(\pmb{X}) = 2X_{1}^2+2cos(\pi X_{2})$, sample size $n=500$, and $\sigma^2=0.1$.  
	}
	\begin{tabular}{ccllllll}
		\hline
		Model& method & minEpan  &Epsilon  & Exp & ZA(1,1.5) & ZA(2,2.5)&ZA(3,3.5)  \\ 
		\hline
		\multirow{4}*{SP}
		& $c=1.0$  & 0.73(0.22) & 1.18(0.46) & 0.91(0.37) & 0.80(0.24) & 0.92(0.29) & 1.18(0.49) \\ 
		& $c=2.0$  & 1.15(0.32) & 2.99(0.75) & 2.54(0.40) & 1.31(0.39) & 1.28(0.36) & 1.38(0.46) \\ 
		& $c=3.0$  & 1.68(0.52) & 4.15(0.91) & 3.75(0.66) & 2.09(0.64) & 1.84(0.60) & 1.89(0.66) \\ 
		& $c=4.0$  & 2.40(0.84) & 5.14(1.16) & 4.79(0.92) & 3.11(0.96) & 2.61(0.87) & 2.62(0.90) \\ 
		\hline
		\multirow{4}*{EXP} 
		& $c=2.5$  & 0.85(0.22) & 1.42(0.46) & 1.34(0.41) & 0.92(0.24) & 1.04(0.31) & 1.25(0.48) \\
		& $c=2.0$  & 0.96(0.29) & 1.93(0.60) & 1.75(0.44) & 1.03(0.30) & 1.12(0.36) & 1.32(0.48) \\  
		& $c=1.5$  & 1.16(0.40) & 2.59(0.75) & 2.31(0.50) & 1.28(0.44) & 1.31(0.51) & 1.45(0.55) \\ 
		& $c=1.0$  & 1.56(0.46) & 3.54(0.84) & 3.22(0.56) & 1.84(0.48) & 1.67(0.48) & 1.82(0.57) \\
		\hline
		\multirow{4}*{INVQ}
		& $c=10.0$  & 0.93(0.28) & 1.62(0.55) & 1.49(0.49) & 1.02(0.31) & 1.07(0.35) & 1.24(0.45) \\ 
		& $c=7.0$  & 0.99(0.31) & 1.95(0.59) & 1.70(0.50) & 1.06(0.33) & 1.10(0.38) & 1.30(0.48) \\ 
		& $c=3.0$  & 1.40(0.47) & 3.03(0.77) & 2.70(0.49) & 1.59(0.48) & 1.53(0.50) & 1.65(0.56) \\ 
		& $c=1.0$  & 2.06(0.74) & 4.51(0.99) & 4.11(0.83) & 2.53(0.84) & 2.27(0.78) & 2.29(0.83) \\ 	
		\hline
	\end{tabular}
	\label{MES_D_equal2}
\end{table}

From Table \ref{MES_D_equal2}, it is evident that the minimum $\mbox{\rm MSE}_{prac}$ increases as $\rho_n(t)$ increases. Bimodal kernel functions that are Lipschitz continuous at $0$ are less efficient at removing the effects of error correlation compared to the kernel function $K_z$, such as the Epsilon-class function and the Exponential function. Comparing the three different sets of $(c_1,c_2)$ pairs, it can be seen that when the correlation function is centered around $0$, ``ZA$(1,1.5)$" performs the best among all five methods. This is plausible because when $c_1$ is smaller, the quantity $\left[\mu(K_z^2)\mu_2(K_z)\right]$ will be smaller, resulting in a smaller $\mbox{\rm MSE}$ value. On the other hand, when the error correlation is strong, a small $c_1$ may not be sufficient to mitigate such correlation effect, leading to a larger $\mbox{\rm MSE}$. Therefore, we observe better performance of ``ZA$(2,2.5)$" or ``ZA$(3,3.5)$" when the correlation function is more distant from $0$, such as the Spherical model with $c=4.0$ or the Inverse quadratic model with $c=1.0$.

\subsection{Estimation of covariance function}\label{Sec_Simu2}

Next, we investigate the performance of $\widehat{\sigma^2}$ and the correlation function estimator $\widehat{\rho}_n(t)$. We compare the three different methods mentioned in section \ref{Sec_Simu1}. Additionally, we use the estimator computed assuming the errors are known as the reference line for comparison. This includes the variance estimator $\widetilde{\sigma^2}=\frac{1}{n}\sum_{i=1}^n\varepsilon_i^2$ and covariance estimator $\widetilde{C}_n(t)$. (denoted as ``Raw").

First, we study the simulation results of $\widehat{\sigma^2}$, which is computed as in equation \eqref{Sec3_VarEst1}. The $\widehat{\varepsilon}_i$ are obtained via $\widehat{\varepsilon}_i = Y_i - \widehat{\mu}_{h^T, K_o}(\mathbf{X}_i)$ with $h^T =\widehat{h}{n^{-\frac{1}{D+8}}}\big/{n^{-\frac{1}{D+4}}}$, where $\widehat{h}$ is the bandwidth chosen from the three different methods in section \ref{Sec_Simu1}. We use
\begin{equation*}
	\mbox{\rm MSE}_{\widehat{\sigma^2}} = \frac{1}{N_T}\sum_{T=1}^{N_T}\left(\widehat{\sigma^2}^{(T)}-\sigma^2\right)^2
\end{equation*}
to measure the performance of $\widehat{\sigma^2}$, where $N_T$ is the number of trials in the simulation and $\widehat{\sigma^2}^{(T)}$ is the variance estimator of $T$-th simulation trial.

\begin{table}[h!]
	\centering
	\caption{$\mbox{\rm MSE}_{\widehat{\sigma^2}}$ ($\times 10^{-5}$) in each simulation scenario from 100 trials. The regression function is $\mu(\X) = 2X_{1}^2+2cos(\pi X_{2})$, sample size $n=500$,  and $\sigma^2=0.1$.  }
	\begin{tabular}{ccllllll}
		\hline
		Model& method & Raw & Epsilon & Exp & ZA(1,1.5) & ZA(2,2.5)&ZA(3,3.5)  \\ 
		\hline
		\multirow{4}*{SP}
		&	$c=1.0$ & 4.60 & 9.89 & 8.69 & 10.23 & 20.59 & 38.79 \\ 
		&	$c=2.0$ & 6.49 & 18.17 & 56.55 & 10.16 & 20.18 & 29.31 \\ 
		&	$c=3.0$ & 10.49 & 97.07 & 141.83 & 24.66 & 18.70 & 23.36 \\ 
		&	$c=4.0$ & 16.44 & 224.37 & 260.00 & 74.60 & 25.27 & 23.65 \\ 
		\hline
		\multirow{4}*{EXP} 
		&$c=2.5$ & 4.52 & 6.56 & 12.86 & 8.34 & 18.91 & 34.88 \\ 
		&$c=2.0$ & 4.82 & 4.79 & 23.70 & 8.86 & 17.80 & 34.81 \\ 
		&$c=1.5$ & 5.53 & 10.90 & 46.87 & 10.53 & 15.66 & 33.48 \\ 
		&$c=1.0$ & 7.79 & 48.20 & 103.36 & 16.54 & 10.78 & 27.65 \\ 
		\hline
		\multirow{4}*{INVQ}
	&	$c=10.0$ & 4.68 & 5.21 & 15.15 & 8.53 & 15.15 & 29.09 \\ 
	&	$c=7.0$ & 4.94 & 5.58 & 23.41 & 7.17 & 14.01 & 34.71 \\ 
	&	$c=3.0$ & 6.28 & 20.95 & 64.69 & 11.51 & 11.14 & 22.65 \\ 
	&	$c=1.0$ & 11.27 & 117.53 & 173.16 & 36.83 & 15.74 & 23.40 \\  		
		\hline
	\end{tabular}
	\label{Sig2_D_equal2}
\end{table}

From Tables \ref{Sig2_D_equal2}, it is observed that when the correlation function is relatively large, using proposed kernel $K_z$ will lead to a smaller $\mbox{\rm MSE}_{\widehat{\sigma^2}}$ compared to the Epsilon-class kernel and Exponential kernel. Additionally, as the correlation function becomes stronger, using $K_z$ that is farther away from 0 can provide a better estimation of $\sigma^2$. However, we can see that if the correlation function is small, the Epsilon-class kernel may provide a better estimation of $\sigma^2$. This suggests that if the correlation function is very weak, using the Epsilon-class kernel may lead to better bandwidth selection for variance estimation. Nonetheless, the purpose of this article is to pursue better performance when the correlation function is not predominantly centered around $0$.

Second, we investigate the performance of the correlation estimator $\widehat{\rho}_n(\cdot)$. We employ the ``Variance-calibration" procedure described in section \ref{Sec_Cov_est} to select the bandwidth $b$, aiming to find the largest $b$ such that $\vert\widehat{C}_n(0)-\widehat{\sigma^2}\vert\leq \delta_n$. In our simulation, we set $\delta_n = 2\times 10^{-4}$.  After determining $b$, we estimate $\widehat{\rho}_n(t)$ via \eqref{Sec3_rho_est} with the boundary kernel in \eqref{Sec3_boundary_kernel}. To evaluate the performance of $\widehat{\rho}_n(t)$, we use the following quantity:
\begin{align*}
	&\quad \mbox{\rm SSE}_{\rm cor}= \sum_{i=1}^{n}\sum_{j=i+1}^{n}\left\{\widehat{\rho}_n(\Vert\X_i-\X_j\Vert)-\rho_n(\Vert\X_i-\X_j\Vert)\right\}^2I\left(\rho_n\left(\Vert\X_i-\X_j\Vert\right)\leq \zeta \right),
\end{align*}   
and $\zeta$ is a positive threshold used to truncate the correlation estimation when $\rho_n\left(\Vert\mathbf{X}_i-\mathbf{X}_j\Vert\right)$ is smaller than $\zeta$. In our simulation, the threshold $\zeta$ is set as 0.02. Basically, $\mbox{\rm SSE}_{\rm cor}$ measures the performance of correlation estimator when the real correlation is greater than $\zeta$.

Table \ref{Rho_est_D2} presents the simulation results of $\mbox{\rm SSE}_{\rm cor}$ for 2-D models respectively. When the correlation is small, such as in the first line of each correlation model in Table \ref{Rho_est_D2}, we observe that the $\mbox{\rm SSE}_{\rm cor}$ values from the ``Raw" method, where $\varepsilon$ is treated as known, are relatively large compared to others. This is not surprising since $\widehat{\varepsilon}$ will be close to $\varepsilon$ when the correlation is small, and the estimation error of covariance will mostly come from the difference between $\widetilde{\rho}_n(t)$ and $\rho_n(t)$. 

On the other hand, when the correlation is large, the $\mbox{\rm SSE}_{\rm cor}$ values from the ``Raw" method are much smaller than others. In this case, the estimation error will be dominated by the error from $\widehat{\rho}_n(t)$. It is observed that using the proposed kernel $K_z$ performs better than the Epsilon-class kernel and the Exponential kernel most of the time. From section \ref{Sec_band}, we know that if the correlation is large, kernel function $K_z$ with a larger $c_1$ value leads to a better choice of bandwidth, hence $\widehat{\varepsilon}_i$ provides a better estimation of $\varepsilon_i$. Thus, we can see that as the correlation function increases, a kernel function that is farther away from 0 will lead to a better estimation of covariance.

\begin{table}[h!]\footnotesize
	\centering
	\caption{Mean and standard deviation (in parentheses) of $\mbox{\rm SSE}_{\rm cor}$ values in each simulation scenario from 100 trials. The regression function is $\mu(\pmb{X}) = 2X_{1}^2+2cos(\pi X_{2})$, sample size $n=500$, $\sigma^2=0.1$. ``SP", ``EXP" and ``INVQ" stand for Spherical model, Exponential model and Inverse quadratic model respectively. }
	\begin{tabular}{ccrrrrrr}
		\hline
		Model& method & Raw  &Epsilon  & Exp & ZA(1,1.5) & ZA(2,2.5)&ZA(3,3.5)  \\ 
		\hline
		\multirow{4}*{SP}
		& $c=1.0$ & 4.68(3.02) & 25.36(20.17) & 3.19(2.19) & 5.96(4.24) & 10.27(6.77) & 16.01(10.57) \\ 
		& $c=2.0$ & 20.12(21.87) & 256.02(68.45) & 183.65(21.81) & 70.90(37.93) & 41.70(25.26) & 31.21(18.11) \\ 
		& $c=3.0$ & 38.98(52.57) & 700.10(58.56) & 654.62(37.61) & 386.78(130.57) & 258.42(113.41) & 181.93(100.54) \\ 
		& $c=4.0$ & 95.76(82.52) & 1489.61(101.99) & 1410.76(83.30) & 1052.23(201.80) & 710.70(224.53) & 494.16(205.60) \\ 
		\hline
		\multirow{4}*{EXP} 
		&	$c=2.5$ & 13.06(6.26) & 82.23(44.72) & 19.30(14.22) & 16.54(10.83) & 20.42(12.21) & 29.72(17.99) \\ 
		&	$c=2.0$ & 13.55(9.42) & 146.80(65.72) & 40.43(18.09) & 19.77(16.82) & 13.59(12.35) & 19.92(27.84) \\ 
		&	$c=1.5$ & 17.72(18.64) & 249.80(87.42) & 121.38(21.71) & 60.93(38.52) & 32.48(26.65) & 21.23(28.10) \\ 
		&	$c=1.0$ & 53.18(44.40) & 545.45(109.27) & 429.50(31.27) & 282.89(85.04) & 172.75(88.18) & 100.63(69.01) \\ 
		\hline
		\multirow{4}*{INVQ}
		&	$c=10.0$ & 20.76(18.84) & 127.91(49.61) & 40.12(18.42) & 30.03(19.40) & 20.62(13.00) & 32.13(26.81) \\ 
		&	$c=7.0$ & 21.18(17.03) & 181.10(70.73) & 64.73(22.97) & 36.88(26.18) & 21.34(20.42) & 26.99(29.41) \\ 
		&	$c=3.0$ & 44.75(47.74) & 391.20(91.25) & 230.44(36.12) & 151.02(62.29) & 90.91(52.40) & 54.99(39.31) \\ 
		&	$c=1.0$ & 289.11(177.85) & 1081.67(112.79) & 977.74(51.45) & 723.42(136.64) & 554.78(148.01) & 409.89(179.55) \\ 
		\hline
	\end{tabular}
	\label{Rho_est_D2}
\end{table}

\section{Application to Cardiovascular Disease Mortality Rates Data set}\label{Sec_real_data_application}
We return to the dataset on \href{https://ghdx.healthdata.org/record/ihme-data/united-states-cardiovascular-disease-mortality-rates-county-1980-2014}{Cardiovascular Disease Mortality Rates}. Figure \ref{figure_app_raw}(a) shows the mortality rates (deaths per 1000 population) for cardiovascular disease from 1064 counties in the Southeastern United States in the year 2014. 
%\begin{figure}[h!]
%	\centering
%	\includegraphics[scale=0.5]{Raw_data_plot.pdf}
%	\caption{Cardiovascular disease mortality rates of 1064 counties in the southeastern United States in the year 2014.}
%	\label{figure_app_raw}
%\end{figure}
In Figure \ref{figure_app_raw}, there is no clear monotonic trend. Instead, several areas show high mortality rates, such as eastern Kentucky, inner Georgia, and the region where Mississippi, Arkansas, and Louisiana intersect. This suggests that a parametric regression might not be suitable. Therefore, we used the bandwidth selection procedure from Section \ref{Sec_band} to fit the regression surface of the mortality rate. We then utilized the residuals and the methods detailed in Section \ref{Sec_Cov_est} to estimate the error covariance function.

\begin{figure}[h!]
	\centering
	\includegraphics[scale=0.4]{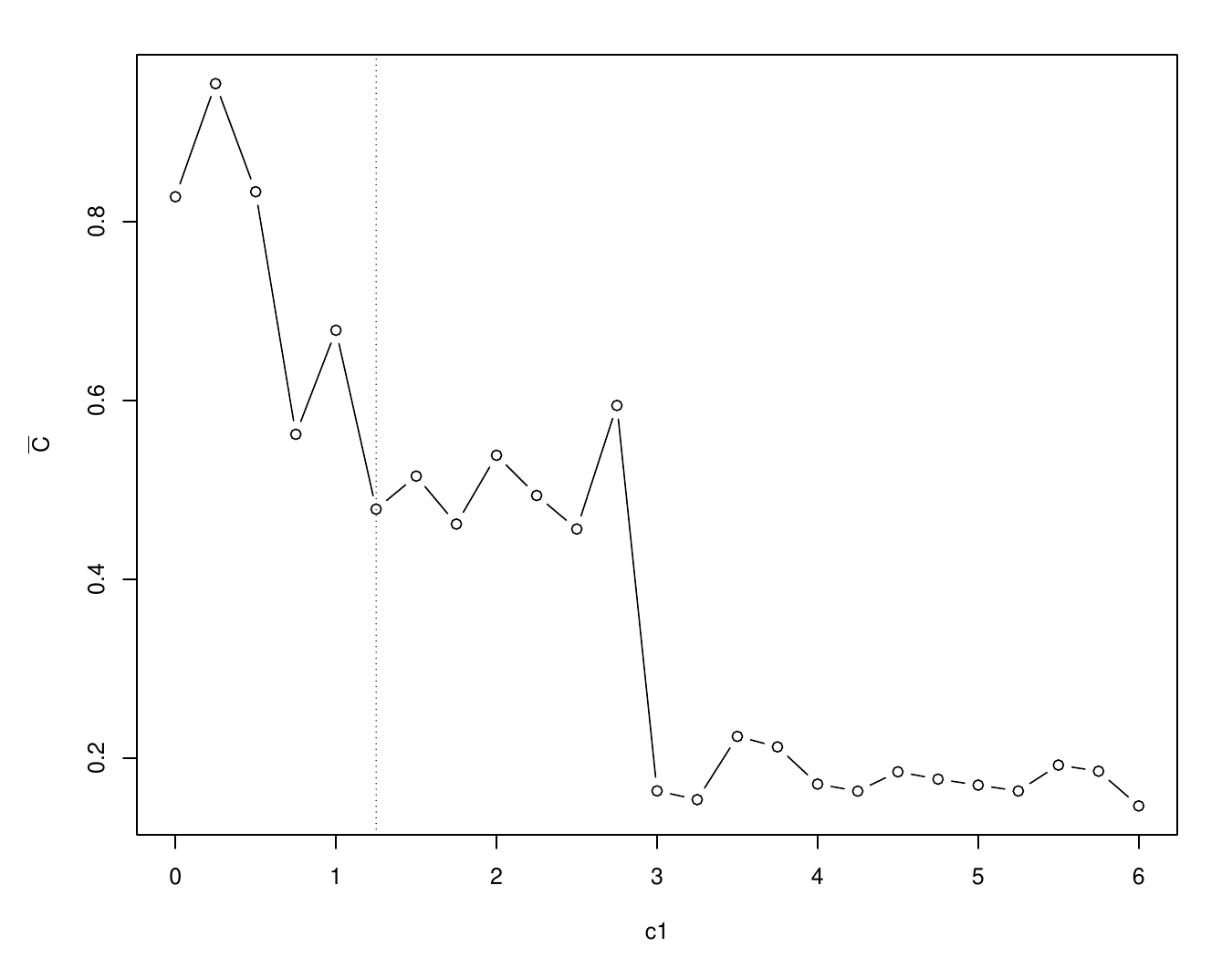}
	\caption{\textit{The visualization of $c_1$ against $\bar{C}$, the plot gives the choice of $c_1=1.25$, which corresponds to the vertical dotted line.}}
	\label{plot_c1_choice}
\end{figure}

For comparison purposes, we applied the three methods described in Section 2.2 to  \ref{Sec_Simu1} for bandwidth selection. In addition, we employed the $C_p$ criterion with the product Epanichikov kernel (referred to as ``Epan"), which does not consider the error correlation at all. The ``elbow method" was applied to seek for an appropriate $c_1$. Figure \ref{plot_c1_choice} present the visualization of $c_1$ against $\bar{C}$, when the candidate set of $c_1$ is $\{0,0.25,\cdots,5.75,6.00\}$ and $c_2=c_1+0.5$. Based on Figure \ref{plot_c1_choice}, we pick $c_1=1.25$ such the $\bar{C}$ values are stable for the first time.

\begin{figure}[h!]
	\centering
	\subfigure[Epan]{\includegraphics[scale=0.4]{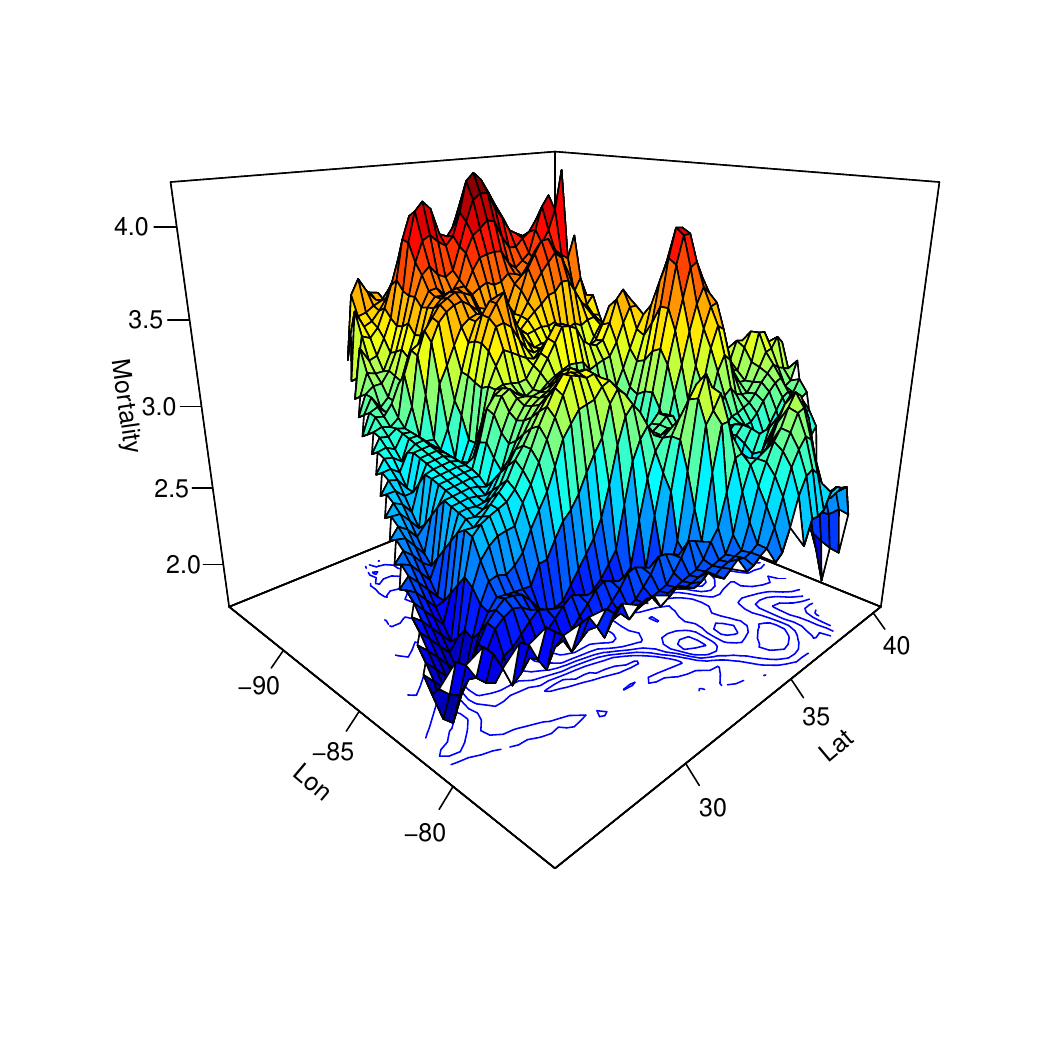}}
	\subfigure[Eps]{\includegraphics[scale=0.4]{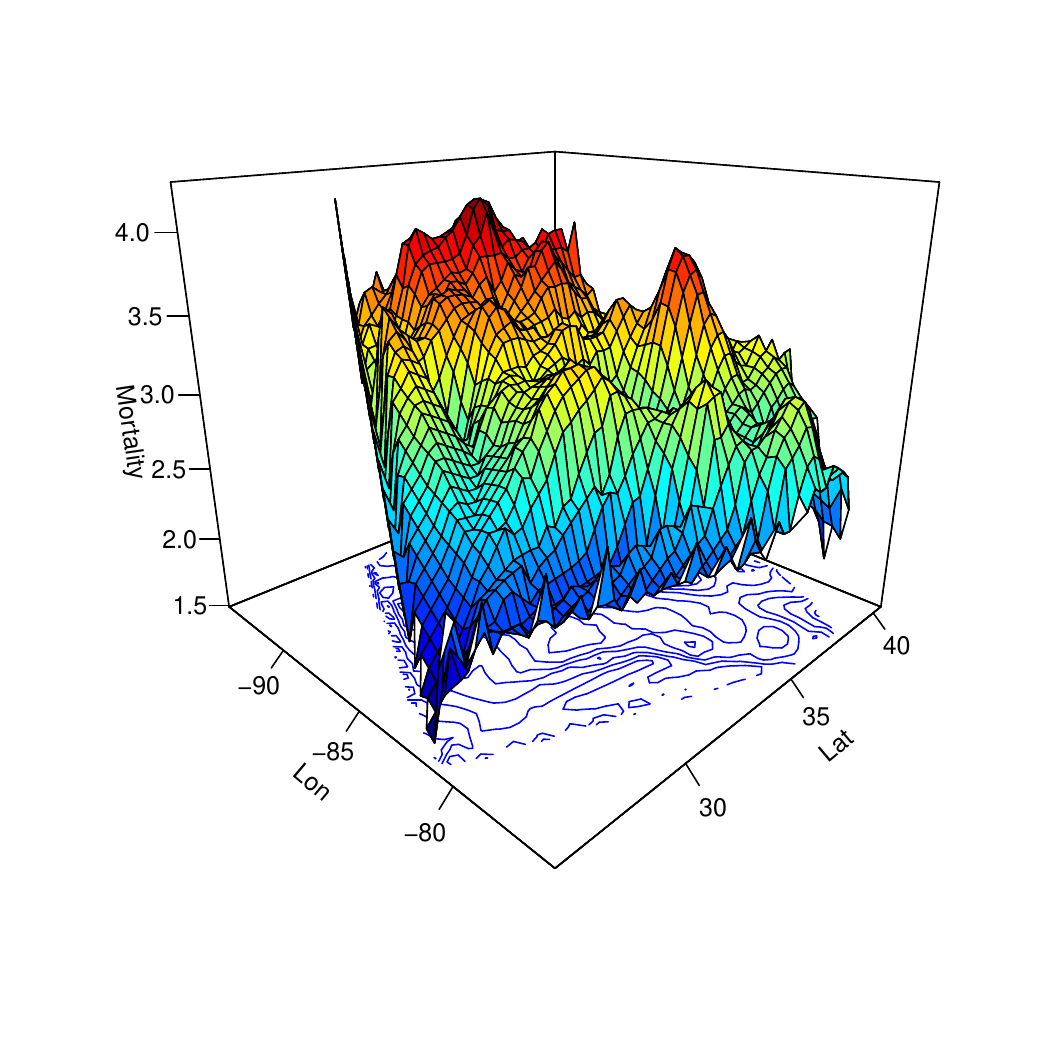}}
	\\
	\subfigure[Exp]{\includegraphics[scale=0.4]{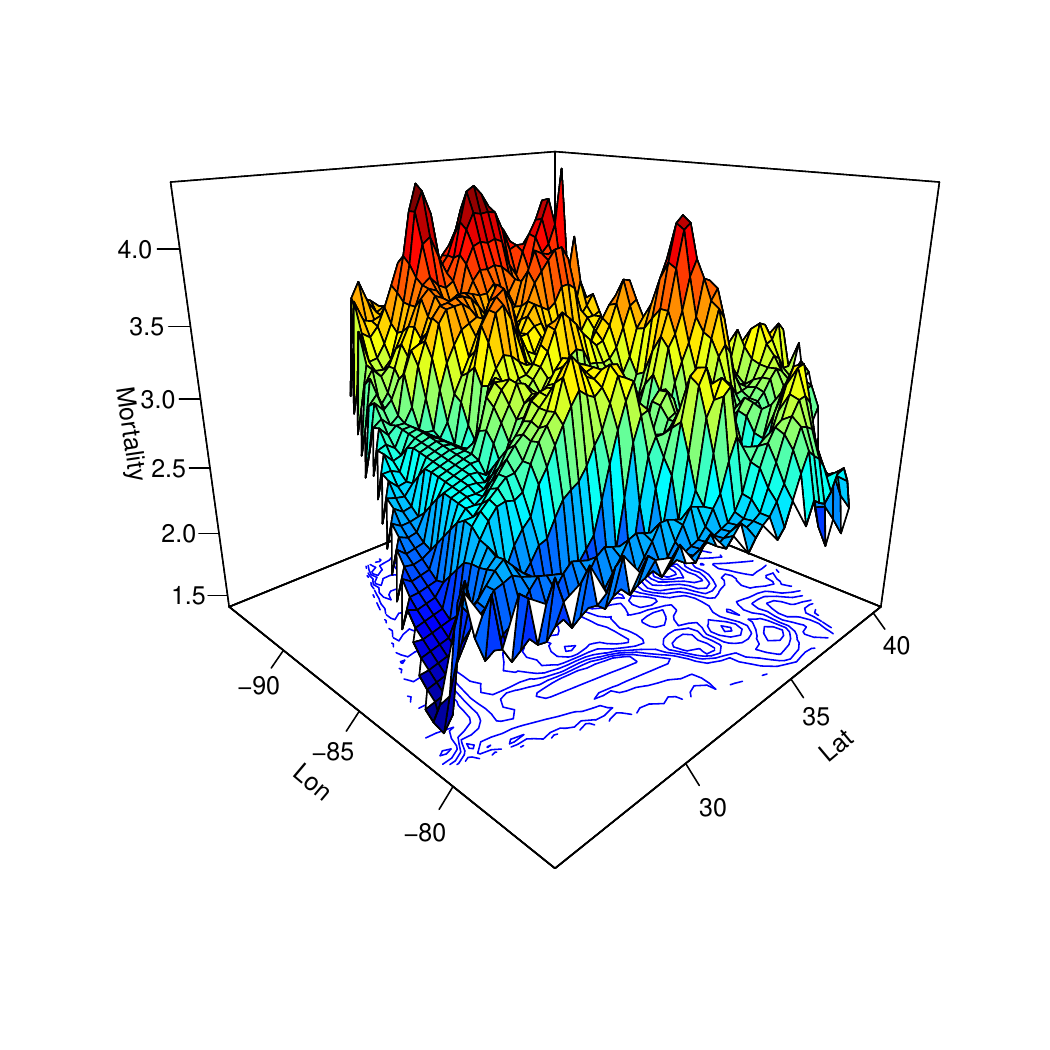}}
	\subfigure[ZA$(1.25,1.75)$]{\includegraphics[scale=0.4]{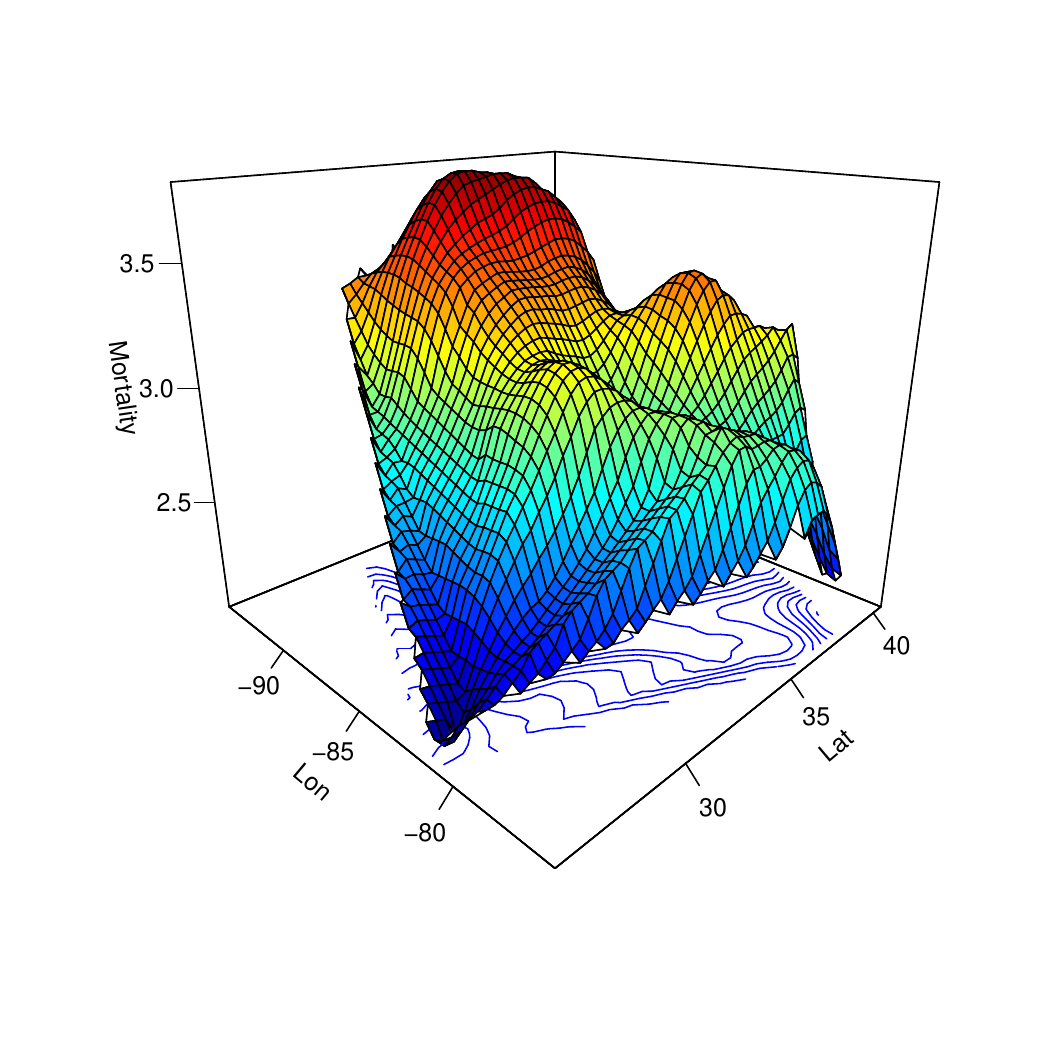}}
	\caption{\textit{(a)Epan, bandwidth selected from using $C_p$ criterion with product Epanichikov kernel; (b)Eps, bandwidth selected from $C_p$ criterion with epsilon-class kernel; (c)Exp, bandwidth selected from exponential kernel with factor method; (d)ZA, bandwidth selected from Zero-apart kernel with factor method. $(c_1,c_2)=(1.0,1.5)$. }}
	\label{figure_app_surface}
\end{figure}

\begin{figure}[h!]
	\centering
	\subfigure[Contour plot]{\includegraphics[scale=0.35]{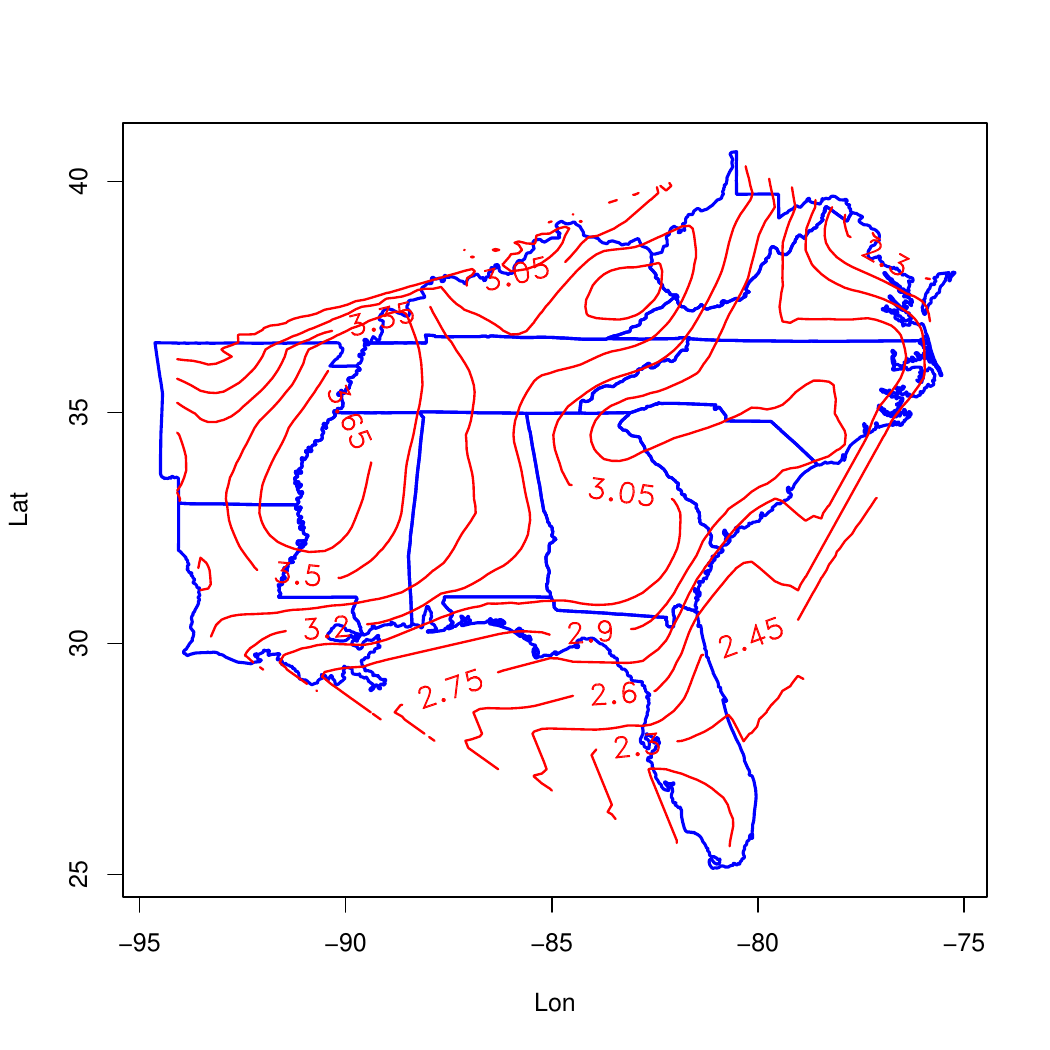}}
	\subfigure[Correlation]{\includegraphics[scale=0.35]{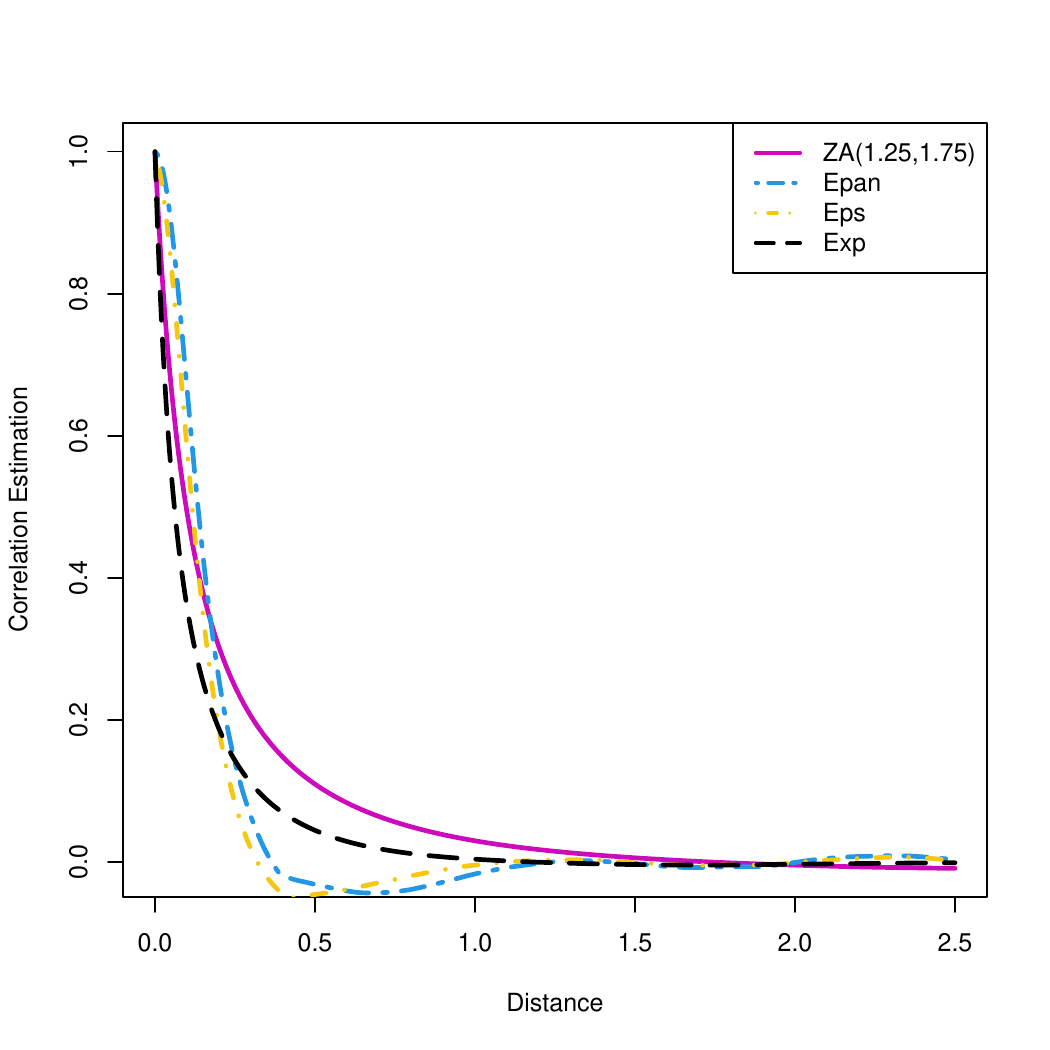}}
	\caption{(a): contour plot of the fitted surface with the application of kernel \eqref{Sec2_kernel}. (b): relations between the estimated error correlation and distance ($\times 10^2$km).}
	\label{figure_plot_results}
\end{figure}

The panel (a) to panel (d) in Figure \ref{figure_app_surface} display the fitted surface plots from four different bandwidth selections. The application of the polynomial kernel \eqref{Sec2_kernel} in panel (d) effectively removes most of the correlation structure, resulting in a smoother surface estimation. When we employ the other three methods, the fitted surface exhibits multiple peaks, which leads to difficulty in interpretation. For better visualization, the contour plot of the panel (d) of Figure \ref{figure_app_surface} is displayed in Figure \ref{figure_plot_results}(a). It is observed that the three areas, including eastern Kentucky, inner Georgia, and the region where Mississippi, Arkansas, and Louisiana intersect, exhibit relatively higher mortality rates compared to the surrounding regions.

Next, we aim to estimate the error correlation function to explore the relationship between the mortality rates and the geological distance of the two counties. We utilized the Haversine method to compute the actual distance between two points instead of the Euclidean distance. Figure \ref{figure_plot_results}(b) illustrates that the estimated correlation function decreases as the distance between two locations increases. Moreover, employing the kernel function \eqref{Sec2_kernel} results in a more intense correlation function, which decreases slower than other estimates. It indicates that Cardiovascular Disease Mortality Rates from two counties that are more than $50$km away may still related to each other. In real-world applications, the true error correlation function is unknown. However, it is evident that the correlation function estimated using bandwidth selected by the naive $C_p$ criterion, without considering any correlation, is inappropriate. Note that the correlation functions estimated from ``Eps" are very close to that from ``Epan". This suggests that the epsilon-class may not behave well for the correlation estimation in this application.

\textit{Remark:}
In addition to the Cardiovascular Disease Mortality Rates data, we extracted Life Expectancy and Colon \& Rectum cancer mortality rates for all counties in the Southeastern United States from the \href{https://ghdx.healthdata.org/record/ihme-data/united-states-adult-life-expectancy-county-1987-2007}{Life Expectancy Data} and \href{https://ghdx.healthdata.org/record/ihme-data/united-states-cancer-mortality-rates-county-1980-2014}{Cancer Mortality Rates data} respectively. We applied our methods to these two datasets as well. Please refer to the supplementary file for details of the additional real data applications. All three real data applications demonstrate that a bimodal kernel will not be sufficient to remove error correlation, resulting in a wiggly surface that is difficult to interpret. Moreover, \href{https://ghdx.healthdata.org/}{the Global Health Data Exchange} also contains a variety of other disease mapping data, and our methods are applicable to many of these data types.

\section{Discussion}\label{Sec_discuss}
In the present article, we introduce a method for selecting the bandwidth parameter to estimate the regression function in the presence of error correlation, and we study the estimation of the error correlation function after estimating the regression function. Simulation studies showed that our proposed method performs well when the correlation function is not predominantly centered around zero. Finally, we demonstrate the application of the proposed method by analyzing Cardiovascular Disease Mortality Rates in the Southeastern United States.

We envision several directions for future research. First, the bandwidth selection method proposed for the error variance estimator, $\widehat{\sigma^2}$ remains heuristic. In the future, we aim to conduct a rigorous investigation to identify a more precise approach for selecting the bandwidth for the estimation of the error covariance function. Second, it would be an interesting problem to develop a statistical test to identify where the error correlation function diminishes to 0. Third, since the Spatial-Temporal data are characterized by variations in both time and location, we plan to extend our method to accommodate spatial-temporal data models of the form $Y_i = \mu\left(t_i,\boldsymbol{X}_i\right) +\varepsilon_i$ with $\varepsilon_i$'s are correlated, where $t_i$ and $\boldsymbol{X}_i$ denote the observed time and location, respectively. Finally, our method still assumes homogeneous error variance, but it is worth extending it to account for heteroskedastic error variance which may fit with more real data applications.

%\backmatter
\bmsection*{Author contributions}
Sisheng Liu worked for the Methodology development, theoretical studies, simulation, real data analysis and writing. Xiao likong worked for the writing and theoretical studies.

\bmsection*{Acknowledgments}
Sisheng Liu is supported by the National Natural Science Foundation of China (Grant 12401346) and the Natural Science Foundation of Hunan Province (Grant 2025JJ60008).

\bmsection*{Financial disclosure}

None reported.

\bmsection*{Conflict of interest}

The authors declare no potential conflict of interests.

\bibliographystyle{unsrtnat}
\bibliography{wileyNJD-AMA} 

\bmsection*{Supporting information}
None.

\appendix

%\section*{Appendix}\label{Sec_Theory}
\setcounter{equation}{0}
\def\theequation{A.\arabic{equation}}
\setcounter{theorem}{0}
\def\thetheorem{A.\arabic{theorem}}
\setcounter{corollary}{0}
\def\thecorollary{A.\arabic{corollary}}
In this section, we present some key theoretical results that lay the foundation of the methods in Section \ref{Sec_band}. 
Recall that in Section \ref{Sec_band}, we present a three-step method to select $h=\widehat{h}(K_o)$ as the final bandwidth parameter in the estimation \eqref{Sec2_local_linear}. Theorem \ref{Sec2_Thm} stated that $\widehat{h}(K_o)$ is asymptotically equal to the optimal bandwidth $h^*$ which minimizes $\MISE$. To prove this, several key results are leading to the theorem. We introduce the mean average square error defined as follows:
\begin{equation*}
	\mbox{\rm MASE}(h,K) =  \E\left[\frac{1}{n}\sum_{i=1}^{n}(\mu(\X_i) - \widehat{\mu}_{h,K}(\X_i))^2\bigg\vert \X\right],
\end{equation*}
where $h$ is the bandwidth and $K$ represents the kernel function. MASE is a mathematical approximation for the integrate MISE. 
Next, we show that with kernel function $K_z$, we can remove the correlation structure.
\begin{theorem}\label{Sec6_Thm6.1}
	Suppose model \eqref{Sec1_model} and assumptions A1-A3 holds, the kernel function $K_z$ satisfies C1 to C3 with non-zero support on an annular regions $c_1<\Vert\pmb{u}\Vert<c_2$ and the bandwidth $h\in (a_1n^{-\frac{\alpha}{D+4}},a_2n^{-\frac{\alpha}{D+4}})$ for some positive numbers $a_1$ and $a_2$, then there exist $c_1$ such that
	\begin{equation*}
		\mbox{\rm MASE}(h,K_z) = \E\left[\RSS(h,K_z)\vert \X\right] - \sigma^2 + o_p\left(\frac{1}{n^\alpha h^D}\right).
	\end{equation*}
\end{theorem}

\begin{corollary}\label{Sec6_Cor6.1}
	Suppose that conditions in Theorem \ref{Sec6_Thm6.1} are satisfied. Then
	\begin{equation}\label{MISE_to_RSS}
		\MISE(h,K_z) = \E[\RSS(h,K_z)\vert \X] - \sigma^2 + o_p\left(\frac{1}{n^\alpha h^D}\right). 
	\end{equation}
\end{corollary}

Corollary \ref{Sec6_Cor6.1} indicates that with the kernel function $K_z$, we can remove the error correlation effect for bandwidth selection. We emphasize that this is not generally true with other kernel functions. For the bimodal kernel with only $K(0)=0$, under assumptions A1 to A3, the result of \eqref{MISE_to_RSS} does not hold. The analysis is present in the S2.2 Remark of Supplementary. 
%Corollary \ref{Sec6_Cor6.1} implies that one can minimize $\E[\RSS(h,K_z)\vert\X]$ for finding the minimization of $\MISE(h,K_z)$ with kernel function $K_z$. However, the variance of $\RSS$ should be bounded since $\RSS(h,K_z)$ is the only quantity that can be obtained in practice.

\begin{theorem}\label{Sec6_Thm6.2}
	Suppose model \eqref{Sec1_model} and assumptions A1-A3 hold, the kernel function $K$ has bounded support and the bandwidth $h\in (a_1n^{-\frac{\alpha}{D+4}},a_2n^{-\frac{\alpha}{D+4}})$ for some positive numbers $a_1$ and $a_2$, then 
	\begin{equation*}
		\var(\RSS(h,K)\vert \X) = O_p(n^{-\frac{4\alpha}{D+4}}).
	\end{equation*} 
\end{theorem}

From Theorem \ref{Sec6_Thm6.2}, we see that the conditional variance of $\RSS(h,K_z)$ is bounded by the order of $n^{-\frac{4\alpha}{D+4}}$. Based on Corollary \ref{Sec6_Cor6.1} and Theorem \ref{Sec6_Thm6.2}, it can be shown that the bandwidth minimize $\RSS(h,K_z)$ will asymptotically minimize $\MISE(h,K_z)$. Next, we establish the $\MISE$ of local linear regression estimator from bias and variance decomposition under the assumptions A1 to A3. 

\begin{theorem}\label{Sec6_Thm6.3}
	Suppose model \eqref{Sec1_model} and assumptions A1-A3 holds, the bandwidth satisfies $h\rightarrow 0$ and $n^\alpha h^D\rightarrow \infty$ as $n\rightarrow\infty$, then
	\begin{equation}\label{Sec6_MISE_form}
		\MISE(h,K) =\left\{
		\begin{aligned}
			&\frac{1}{4}h^4\Delta_f^2 \mu_2(K)^2 + \frac{\sigma^2\left(C_\rho+m(\mathcal{X})\right)}{nh^D}\mu(K^2)  + o_p\left(h^4+\frac{1}{nh^D}\right),  \alpha=1,\\
			&\frac{1}{4}h^4\Delta_f^2 \mu_2(K)^2 + \frac{\sigma^2C_\rho}{n^\alpha h^D}\mu(K^2) + o_p\left(h^4+\frac{1}{n^\alpha h^D}\right),  0<\alpha<1.
		\end{aligned}
		\right. 
	\end{equation}
\end{theorem}

From Theorem \ref{Sec6_Thm6.3}, the asymptotic optimal bandwidth for a given kernel function $K$ can be derived directly by minimizing the leading term of MISE in \eqref{Sec6_MISE_form}.

\begin{corollary}\label{Sec6_Cor6.2}
	Suppose that conditions in Theorem \ref{Sec6_Thm6.3} are satisfied, the bandwidth minimize leading term of $\MISE(h,K)$ is 
	\begin{equation*}\label{Sec6_bandwidth_form}
		h_{opt}(K) =\left\{
		\begin{aligned}
			&\left(\frac{4\sigma^2(C_\rho+m(\Omega))}{\Delta_f^2}\frac{\mu(K^2)}{\mu_2(K)^2}\right)^{\frac{1}{D+4}}n^{-\frac{1}{D+4}}, \quad \alpha=1,\\
			&\left(\frac{4\sigma^2C_\rho}{\Delta_f^2}\frac{\mu(K^2)}{\mu_2(K)^2}\right)^{\frac{1}{D+4}}n^{-\frac{\alpha}{D+4}}, \quad  0<\alpha<1.
		\end{aligned}
		\right.
	\end{equation*}
\end{corollary}
Suppose $h_{opt}(K_o)$ and $h_{opt}(K_z)$ are the bandwidth from \eqref{Sec6_bandwidth_form} with kernel function $K_o$ and $K_z$ respectively, then the factor relation gives 
\begin{equation*}\label{bandwidth_fac_form_alpha_1}
	h_{opt}(K_o) = h_{opt}(K_z)\left(\frac{\mu(K_o^2)\mu_2(K_z)^2}{\mu_2(K_o)^2\mu(K_z^2)}\right)^{\frac{1}{D+4}}.
\end{equation*}

\nocite{*}% Show all bib entries - both cited and uncited; comment this line to view only cited bib entries;

\end{document}